\documentclass[aps,twocolumn]{revtex4}
\makeatletter

\newcommand{\Rmnum}[1]{\expandafter\@slowromancap\romannumeral #1@}
\makeatother
\usepackage{graphicx}
\usepackage{dcolumn}
\usepackage{bm}
\usepackage{CJK}
\usepackage{amsfonts}
\usepackage{psfrag}
\usepackage{wrapfig}
\usepackage{subfigure}
\usepackage{makeidx}
\usepackage{multirow}
\usepackage{epsf}
\usepackage{amsmath}
\usepackage{cases}
\usepackage{amssymb}
\usepackage{soul}
\usepackage[version=3]{mhchem}

\usepackage[colorlinks,linkcolor=red,anchorcolor=red,citecolor=blue,urlcolor=blue]{hyperref}

\begin{document}
\title{Non-degenerate Kuznetsov-Ma solitons of Manakov equations and their physical spectra}
\author{Wen-Juan Che$^{1}$}
\author{Shao-Chun Chen$^{1}$}
\author{Chong Liu$^{1,2,3,4}$}\email{chongliu@nwu.edu.cn}
\author{Li-Chen Zhao$^{1,3,4}$}\email{zhaolichen3@nwu.edu.cn}
\author{Nail Akhmediev$^{2}$}\email{Nail.Akhmediev@anu.edu.au}
\address{$^1$School of Physics, Northwest University, Xi'an 710127, China}
\address{$^2$Optical Sciences Group, Department of Fundamental and Theoretical  Physics, Research School of Physics, The Australian National University, Canberra, ACT 2600, Australia}
\address{$^3$Shaanxi Key Laboratory for Theoretical Physics Frontiers, Xi'an 710127, China}
\address{$^4$NSFC-SPTP Peng Huanwu Center for Fundamental Theory, Xi'an 710127, China}
\begin{abstract}
We study the dynamics of Kuznetsov-Ma solitons (KMS) in the framework of vector nonlinear Schr\"odinger (Manakov) equations. Exact multi-parameter family of solutions for such KMSs is derived. This family of solutions includes the known results as well as the previously unknown solutions in the form of the non-degenerate KMSs.
We present the existence diagram of such KMSs that follows from the exact solutions.
These non-degenerate KMSs are formed by nonlinear superposition of two fundamental KMSs that have the same propagation period but different eigenvalues. We present
the amplitude profiles of new solutions, their exact physical spectra, their link to ordinary vector solitons and offer easy ways of their excitation using numerical simulations.
\end{abstract}

\maketitle
\section{Introduction}\label{sec1}

Oscillating localised structures in a wide variety of conservative and dissipative systems known as `breathers' have attracted considerable
interest in the last decades \cite{Book97,Book05,Book08,DB1,DB2,Lattice,PT}. They are known in optics \cite{Dudley1}, hydrodynamics \cite{Dudley2}, Bose-Einstein condensates \cite{Lattice}, micromechanical arrays \cite{Sato}, and in the cavity optomechanics \cite{Wu}.
Breathers play crucial role in understanding various nonlinear coherent phenomena, including modulation instability \cite{MI-AB,Liu2021}, rogue wave events \cite{RW}, Fermi-Pasta-Ulam recurrence \cite{FPU1}, supercontinuum generation \cite{SCG}, and even turbulence \cite{Crespo}.

In conservative integrable systems governed by the scalar nonlinear Schr\"odinger equation (NLSE), fundamental (first order) breathers can be presented in the form of multi-parameter family of solutions that are periodic both in time and in space \cite{TMP1987,MC}. Comprehensive analysis of all physical effects described by this family is difficult due to the presence of free parameters in the solution and large variety of possibilities \cite{MC}. Subdividing the whole family into particular cases simplifies the task. Among the particular cases of this general family we can select the class of breathers on a plane wave background that are periodic in transverse direction \cite{TMP1987,MC}. These are known as Akhmediev breathers. Another class of solutions is solitons on a constant background. Due to the beating between the soliton and the background, these solitons are periodic along the propagation direction. These are known as Kuznetsov-Ma solitons (KMS) \cite{KMa,KMb}. As these solitons are periodic, sometimes they are also dubbed as KM breathers. Periods in each of these subfamilies of solutions are variable parameters. When the period in either space or time becomes large, the common limit of each of these subfamilies is the Peregrine soliton. It has infinite breathing period both in time and in space thus describing an isolated event such as a rogue wave. Recent results \cite{PS1,PS2} reveal a universal role that the Peregrine soliton plays in complex dynamics of multi-soliton evolution.

KMS is oscillating due to the coherent interaction with a constant background \cite{Wabnitz,KM1}.
When the amplitude of the background tends to zero, the period of oscillations increases and in the zero limit the KMS turns into an ordinary bright soliton \cite{Wabnitz}.
The periodic evolution in propagation of the KMS has been observed experimentally both in fiber optics \cite{KMO} and in hydrodynamics \cite{KMH}. KMSs should not be confused with pulsating solitons in dissipative optical systems \cite{Lucas,Peng} where the physical reason for soliton oscillations is different. Oscillations can also appear as a result of beating between several solitons in higher-order solutions
\cite{HOS1,HOS2,HOS3}.

The single NLSE describes nonlinear dynamics of a  scalar wave field.
On the other hand, nonlinear interaction of two or more coupled wave components is common in physics. Such interaction is important in optical fibres \cite{OF}, in Bose-Einstein condensates \cite{BEC} and in multi-directional wave dynamics in the open ocean known as `crossing seas' \cite{F}. The mathematical model that describes the interaction of two wave components is commonly based on Manakov equations \cite{VB0,VB1,VB2,VB3,VB4,VB5,VB6,VB7,VB8,VB9,VB10,VB11,Vobservation1,Vobservation2}. The interaction between the two wave components
 makes the wave dynamics more complex. One example is the presence of unusual dark and four-petal structures in such systems \cite{VB10,VB11}.
Another example is dark breathers with infinite period (dark rogue waves). The latter have been observed experimentally in fiber optics \cite{Vobservation1, Vobservation2}.
Even when considering common soliton solutions, Manakov equations admit qualitatively new types of formations such as non-degenerate solitons \cite{NDS1,NDS2,NDS3,NDS4}.

In this work, we present theoretical and numerical studies of KMS dynamics in the model governed by the Manakov equations. We derived a  family of multi-parameter vector KMS solutions. We show that this family contains a non-degenerate family that has no analogs in the case of scalar KMSs. We analyse their amplitude profiles, their physical spectra and using numerical simulations suggest an easy way of excitation of these solutions. Just as in the scalar case, solitons are the limiting cases of KMSs in the vector model as well. However, finding the link between the KMSs and ordinary solitons in the vector case has not been addressed so far.
Here, we fill this gap in the existing knowledge and found the limit of vector KMSs when they are converted to the non-degenerate vector solitons.

\section{Manakov equations, their KMS solutions and the corresponding symmetries}\label{sec2}

The Manakov equations in dimensionless form are given by \cite{MM}:
\begin{equation}\label{eq1}
\begin{split}
i\frac{\partial\psi^{(1)}}{\partial t}+\frac{1}{2}\frac{\partial^2\psi^{(1)}}{\partial x^2}+(|\psi^{(1)}|^2+|\psi^{(2)}|^2)\psi^{(1)}&=0,\\
i\frac{\partial\psi^{(2)}}{\partial t}+\frac{1}{2}\frac{\partial^2\psi^{(2)}}{\partial x^2}+(|\psi^{(1)}|^2+|\psi^{(2)}|^2)\psi^{(2)}&=0,
\end{split}
\end{equation}
where $\psi^{(1)}(t,x)$, $\psi^{(2)}(t,x)$ are the two nonlinearly coupled components of the vector wave field. The physical meaning of variables $x$ and $t$ depends on a particular physical problem of interest. In optics, $t$ is commonly a normalised distance along the fibre while $x$ is the normalised time in a frame moving with the group velocity \cite{OF}. In the case of Bose-Einstein condensates, $t$ is time while $x$ is the spatial coordinate \cite{BEC}.
The choice of signs for the dispersion and nonlinear terms in Eqs. (\ref{eq1}) corresponds (in optics) to the self-focusing effect and anomalous dispersion regime.

A fundamental (first-order) solution for vector KMS [i.e., $\psi^{(j)}_1(t,x)$] in compact form obtained using a Darboux transformation scheme is given by:
\begin{equation}
\psi^{(j)}_1(t,x)=\rho_j\psi_{0}^{(j)}(t,x) \psi_{km}^{(j)}(t,x),\label{eqkmb}
\end{equation}
where $\psi_{0}^{(j)}$ is the background vector plane wave
\begin{equation}
\psi_{0}^{(j)}= a_j \exp{\left\{i \left[\beta_jx + \left(a_1^2+a_2^2- \frac{1}{2}\beta_j^2 \right)t \right] \right\}}
\label{eqpw}
\end{equation}
with $a_j$ being the two real amplitudes and $\beta_j$ the wavenumbers of the background waves, respectively, while
\begin{eqnarray}
\rho^{(j)}&=&\frac{\tilde{\bm\chi}^*+\beta_j}{\tilde{\bm\chi}+\beta_j}\sqrt{\frac{(\bm\chi^*+\beta_j)(\tilde{\bm\chi}^*+\beta_j)}{(\bm\chi+\beta_j)
(\tilde{\bm\chi}+\beta_j)}},\\
\psi_{km}^{(j)}&=&\frac{\varpi\cosh({\bm{\Gamma}+i\delta_j})
+\cos{(\bm{\Omega}+i\gamma_j)}}{\varpi\cosh{\bm{\Gamma}}+\cos{\bm{\Omega}}},\label{eqkmbp}
\end{eqnarray}
where
\begin{eqnarray}\label{gm}
&\bm{\Gamma}=\alpha (x+\bm\chi_{r}t)+\frac{1}{2}\log\left(\frac{\alpha+\bm\chi_{i}}{\bm\chi_{i}}\right),\\  \label{om}
&\bm{\Omega}=\Omega t=\alpha\left(\frac{\alpha}{2}+\bm\chi_{i}\right)t.
\end{eqnarray}
and
\begin{equation}
\tilde{\bm\chi}=\bm\chi+i\alpha
\end{equation}
where $\alpha$ is a real parameter. Without loss of generality, we set $\alpha\geq0$.
Subscripts $r$ and $i$ in (\ref{gm}) and (\ref{om}) denote the real and imaginary parts of the complex parameter $\bm\chi$, respectively. The latter is the eigenvalue of the Manakov system (\ref{eq1})
which obeys the relation:
\begin{equation}
1+\sum_{j=1}^2\frac{a_j^2}{(\bm\chi-\beta_j)(\tilde{\bm\chi}-\beta_j)}=0.\label{eqchi}
\end{equation}
The relation between the eigenvalue and the spectral parameter of the associated Lax pair is given by:
\begin{equation}\label{Eqlambda}
\lambda=\bm\chi-\sum_{j=1}^2\frac{a_j^2}{\bm\chi+\beta_j}.
\end{equation}
Other parameters in Eq. (\ref{eqkmbp}) are:
\begin{eqnarray}
\delta_j&=&\arg[{(\bm\chi^*+\beta_j)(\bm\chi+{i\alpha}+\beta_j)}],\\
\gamma_j&=&-\frac{1}{2}\log\left[\frac{(\bm\chi^*-{i\alpha}+\beta_j)(\bm\chi+{i\alpha}+\beta_j)}{(\bm\chi^*+\beta_j)(\bm\chi+\beta_j)}\right],\\
\varpi&=&\frac{\alpha+2\bm\chi_{i}}{2\alpha+2\bm\chi_{i}}\sqrt{\frac{\alpha+\bm\chi_{i}}{\bm\chi_{i}}}.
\end{eqnarray}
From here, one can readily confirm that $|\rho^{(j)}|=1$. The solution (\ref{eqkmb}) depends on the background parameters ($a_j$, $\beta_j$) and the real parameter $\alpha$.

From a physical perspective, an important parameter is the relative wave number $\beta_1-\beta_2$, since it cannot be eliminated through Galilean transformation. Indeed, when $\beta_1=\beta_2$, for any eigenvalue given by Eq. (\ref{eqchi}), $\psi^{(1)}_1$ is merely proportional to $\psi^{(2)}_1$, i.e., $\psi^{(1)}_1/\psi^{(2)}_1=a_1/a_2$.
The solution (\ref{eqkmb}) contains the trivial vector generalisation of the scalar KMS solution which has been found in \cite{Priya}.
Our solution is far from being a simple rotation on a [$\psi^1,\psi^2$]-plane. As it will be shown below, it has nontrivial properties of vector KMS once $\beta_1\neq\beta_2$.
Without loss of generality, we can set $\beta_1=-\beta_2=\beta\neq0$.

Physically, the solution (\ref{eqkmb}) describes solitons located on top of plane wave backgrounds (\ref{eqpw}). Such solitons are localized in $x$ (the width is $\sim 1/\alpha$) and they propagate along $t$ with the group velocity $V_g=-\bm\chi_r$. Due to the beating with the background, the amplitude of the soliton oscillates periodically along the $t$-axis. In the limiting case of the infinite period, which implies $\alpha\rightarrow0$, the solution (\ref{eqkmb}) is transformed into the vector rogue wave.

The solution (\ref{eqkmb}) has two symmetries.
The first one is the symmetry of the solution (\ref{eqkmb}) relative
to the sign change of $\beta$ and simultaneous exchange of the wave component.
When the background amplitudes are equal $a_1=a_2=a$, this means:
\begin{eqnarray}
\psi^{(1)}_1(\beta)=\psi^{(2)}_1(-\beta),~~~\beta\neq0.\label{eqsymm1}
\end{eqnarray}

The second symmetry of the vector solution is more complicated. Namely, if $\bm\chi_i\Rightarrow-\bm\chi_i-\alpha$, then
\begin{eqnarray}
\psi^{(j)}_1[(x,t);\bm\chi_i]=\psi^{(j)}_1[(x',t');-\bm\chi_i-\alpha]e^{i\gamma},\label{eqsymm2}
\end{eqnarray}
where $x'=x+\Delta x$, $t'=t+\Delta t$, with $\Delta x$ and $\Delta t$ being fixed constant shifts along the $x$ and $t$ axes respectively and $\gamma$ is a constant phase. The shifts are given by
\begin{eqnarray}
&&\Delta x=-\frac{1}{2\alpha}\left[\frac{4\pi\bm\chi_r}{\alpha+2\bm\chi_i}+\log \left(\frac{\alpha+\bm\chi_i}{\bm\chi_i} \right)\right].\\
&&\Delta t=\frac{4\pi}{\alpha^2+2\alpha \bm\chi_{i}},~~\gamma=2\arg{(\rho)}.
\end{eqnarray}
The symmetry (\ref{eqsymm2}) means that the vector KMSs have periodic amplitude profiles:
\begin{eqnarray}
\left|\psi^{(j)}_1[(x,t);\bm\chi_i]\right|=\left|\psi^{(j)}_1[(x',t');-\bm\chi_i-\alpha]\right|.\label{eqsymm2-1}
\end{eqnarray}
The symmetries (\ref{eqsymm1}) and (\ref{eqsymm2}) serve as a basis for revealing the richness of KMS properties found below.

\begin{figure}[htbp]
\centering
\includegraphics[width=84mm]{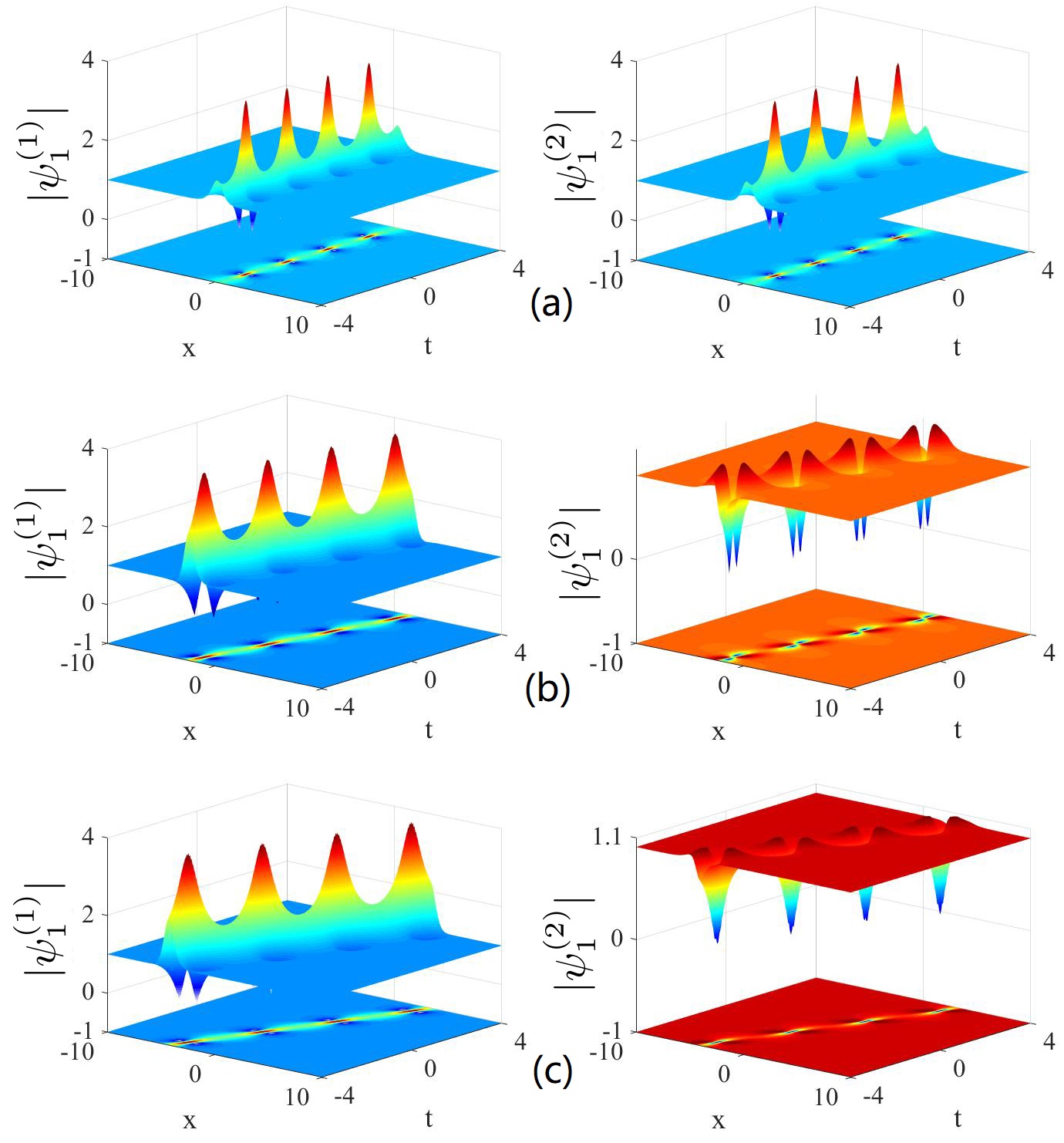}
\caption{Amplitude profiles of the two components of the vector KMS $|\psi^{(j)}_1(t,x)|$ for three different relative background wavenumbers producing qualitatively different wave patterns: (a) $\beta = 0.3$, (b) $\beta = 0.6$, and (c) $\beta = 1$. Other parameters are $a=1$, and $\alpha=2$.}\label{Fig1}
\label{f1}
\end{figure}

The form of the KMS solution (\ref{eqkmb}) has an important advantage. It can be analysed using the  Hessian matrix theory \cite{VB10,VB11}.
In order to do that, we introduce the derivatives $|\psi_{km}^{(j)}|_{\bm\Gamma}$ and $|\psi_{km}^{(j)}|_{\bm\Omega}$.
Then, the zero derivatives $|\psi_{km}^{(j)}|_{\bm\Gamma} = 0$ and $|\psi_{km}^{(j)}|_{\bm\Omega}=0$ define the special points in each cell of $t$-periodic pattern of the KMS. One of them (at the centre) is given by $(\bm\Omega, \bm\Gamma) = (0, \pi)$. The type of these points can be revealed using the Hessian matrix:
\begin{eqnarray}
H^{(j)}=
\left(
\begin{array}{cc}
|\psi_{km}^{(j)}|_{\bm\Gamma\bm\Gamma}&|\psi_{km}^{(j)}|_{\bm\Gamma\bm\Omega}\\
|\psi_{km}^{(j)}|_{\bm\Omega\bm\Gamma}&|\psi_{km}^{(j)}|_{\bm\Omega\bm\Omega} \\
\end{array}
\right).\label{EqH}
\end{eqnarray}
Three distinctive cases which correspond to three different types of KMS can be identified from this analysis. We call them bright, four-petal, and dark solitons. They are shown in
Fig. \ref{f1}. Namely, \\
(1) When $\det(H^{(j)})>0$, and $|\psi_{km}^{(j)}|^2_{\bm{\Gamma}\bm{\Gamma}}<0$, the Hessian is a negative definite matrix.
This implies that the special point is a maximum.  The two components $\psi^{(j)}_1(t,x)$ of the solution in this case have classical `bright' structure. This case is shown  in Fig. \ref{f1}(a). The point of maximal compression in the periodic soliton evolution has a high bump and two small dips at each side of it.\\
(2) When $\det(H^{(j)})<0$, the Hessian is an indefinite matrix.
The centre of each period in this case is a saddle point. The two components of the soliton profile are shown in Fig.\ref{f1} (b). Here, the pattern of the second component $\psi^{(2)}_1(t,x)$ can be called `four-petal' structure.
Namely, each period has two bumps and two dips symmetrically located around the centre. \\
(3) When $\det(H^{(j)})>0$, and $|\psi_{km}^{(j)}|^2_{\bm{\Gamma}\bm{\Gamma}}>0$, the Hessian is a positive definite matrix.
In this case, the second component $\psi^{(2)}_1(t,x)$ is a periodic repetition of dark structures as can be seen in Fig.\ref{f1} (c). The central point in each cell is a minimum. It is surrounded by two small bumps on the sides.

\section{Exact analytic spectra of the KMS}\label{sec3}
Commonly measured characteristics of solitons and breathers are their physical spectra. They are often measured experimentally in optics and hydrodynamics.
One example is the Akhmediev breathers (AB). The AB spectra can be calculated in analytic form \cite{MI-AB}. These spectra are discrete and have an infinite number of sidebands decaying as geometric progression \cite{MI-AB}. Recent experimental observation of more than ten spectral sidebands in an optical fiber
\cite{SD} confirmed the theoretical predictions.
In contrast to the AB which are periodic in transverse variable and therefore have discrete spectra, the spectra of the KMS are continuous.
They can be calculated using the Fourier transform:
\begin{eqnarray}\label{FT}
A_\omega^{(j)}(\omega, t)=\frac{1}{2\pi}\int_{-\infty}^{\infty}\psi^{(j)}(t,x)e^{-i \omega x}dx.
\end{eqnarray}
However, finding the exact analytic KMS spectra is far from being a trivial task due to the symmetry breaking of the Manakov system.
In our previous works \cite{Liu2021,VB11}, we gave some examples of asymmetric discrete spectra in analytic form. Here, we present calculations of the exact analytic continuous spectra of the vector KMS (\ref{eqkmb}).

Let us first rewrite the solution (\ref{eqkmb}) in the form:
\begin{eqnarray}
\psi^{(j)}=\psi^{(j)}_0\rho^{(j)}\left(1+\psi_a(x,t)\right),
\end{eqnarray}
where the new function $\psi_a(x,t)$ is given by
\begin{eqnarray}
\psi_a(x,t)=\frac{\mathcal{B}_1 (t)+\mathcal{B}_2 (t) e^{\alpha x}+\mathcal{B}_3 (t)e^{-\alpha x}}
{\mathcal{D}_1 (t)+\mathcal{D}_2 (t)e^{ \alpha x}+\mathcal{D}_3 (t)e^{-\alpha x}},
\end{eqnarray}
with
\begin{eqnarray}
\mathcal{B}_1 (t)&=&e^{i\bm\Omega}+e^{-i\bm\Omega}-e^{\gamma_j-i\bm\Omega}-e^{-\gamma_j-i\bm\Omega},\nonumber\\
\mathcal{B}_2 (t)&=&\varpi(e^{\bm{\Gamma}-\alpha x}-e^{\bm{\Gamma}-\alpha x+i\delta_j}),\nonumber\\
\mathcal{B}_3 (t)&=&\varpi(e^{-(\bm{\Gamma}-\alpha x)}-e^{-(\bm{\Gamma}-\alpha x)+i\delta_j}),\nonumber\\
\mathcal{D}_1 (t)&=&e^{\bm{\Gamma}-\alpha x}-e^{-(\bm{\Gamma}-\alpha x)},\nonumber\\
\mathcal{D}_2 (t)&=&\varpi e^{\bm{\Gamma}-\alpha x},~~\mathcal{D}_3 (t)=\varpi e^{-(\bm{\Gamma}-\alpha x)}.\nonumber
\end{eqnarray}
The essential part of the integral (\ref{FT}) is the Dirac delta function $\delta(\omega-\beta_j)$ caused by the presence of the background $\psi^{(j)}_0$. We shall omit it in further calculations. The nontrivial part of the integral (\ref{FT}) is:
\begin{eqnarray}\label{FTI}
\mathcal{I}=\frac{1}{2\pi}\int_{-\infty}^{\infty}\psi_a(x,t) e^{-i \omega x} dx.
\end{eqnarray}
The integral in (\ref{FTI}) can be calculated analytically using a residue theorem. Namely,
$\mathcal{I}=2\pi i\mathcal{R}$, where $\mathcal{R}$ is the residue of the corresponding singularity of $\psi_a$ in $x$.
The function $\psi_a$ has two singularities at the points $x_1$ and $x_2$ which are given by
\begin{eqnarray}
x_1&=&\frac{1}{\alpha}\log\left[\frac{1}{2\mathcal{D}_2}(-\mathcal{D}_1-\sqrt{\mathcal{D}_1^2-4\mathcal{D}_2\mathcal{D}_3})\right],\\
x_2&=&\frac{1}{\alpha}\log\left[\frac{1}{2\mathcal{D}_2}(-\mathcal{D}_1+\sqrt{\mathcal{D}_1^2-4\mathcal{D}_2\mathcal{D}_3})\right].
\end{eqnarray}
The explicit expressions for the corresponding residues $\mathcal{R}_{x_1}$ and $\mathcal{R}_{x_2}$ at $x=x_1$ and $x=x_2$, are given by:
\begin{eqnarray}
\mathcal{R}_{x_1}&=&\frac{1}{2\pi\alpha\mathcal{D}_2\mathcal{X}}\left(\frac{\mathcal{H}+\mathcal{P}}
{\mathcal{D}_1+\mathcal{X}}\right)e^{-i\omega x_1},~~\\
\mathcal{R}_{x_2}&=&\frac{1}{2\pi\alpha\mathcal{D}_2\mathcal{X}}\left(\frac{\mathcal{H}-\mathcal{P}}
{\mathcal{D}_1-\mathcal{X}}\right)e^{-i\omega x_2},
\end{eqnarray}
where
\begin{eqnarray} \nonumber
\mathcal{X}&=&\sqrt{\mathcal{D}_1^2-4\mathcal{D}_2\mathcal{D}_3}, \\
\mathcal{H}&=&(\mathcal{D}_1\mathcal{B}_2-\mathcal{D}_2\mathcal{B}_1)\mathcal{X},\nonumber  ~~~~~~~ \mbox{and} \\
\mathcal{P}&=&\mathcal{D}_1(\mathcal{D}_1\mathcal{B}_2-\mathcal{D}_2\mathcal{B}_1)+2\mathcal{D}_2 (\mathcal{D}_2\mathcal{B}_3-\mathcal{D}_3\mathcal{B}_2).\nonumber
\end{eqnarray}

The point $x_1$ is located on the lower complex plane while the point $x_2$ is on the upper complex plane. This means that $\mathcal{I}=2\pi i\mathcal{R}_{x_1}$ when $\omega>0$ while $\mathcal{I}=2\pi i\mathcal{R}_{x_2}$ when $\omega<0$.
Thus, the exact analytic expressions of the KMS spectra can be written as
\begin{eqnarray}\label{FT-KMB}
\begin{split}
A_\omega^{(j)}(\omega, t)&=&
i\frac{(\mathcal{H}+\mathcal{P})}
{\alpha\mathcal{D}_2\mathcal{X}(\mathcal{D}_1+\mathcal{X}^2)}e^{-i\omega x_1},&~\omega>0,\\
A_\omega^{(j)}(\omega, t)&=&
i\frac{(\mathcal{H}-\mathcal{P})}
{\alpha\mathcal{D}_2\mathcal{X}(\mathcal{D}_1-\mathcal{X}^2)}e^{-i\omega x_2},&~\omega<0.\\
\end{split}
\end{eqnarray}

\begin{figure}[htbp]
\centering
\includegraphics[width=84mm]{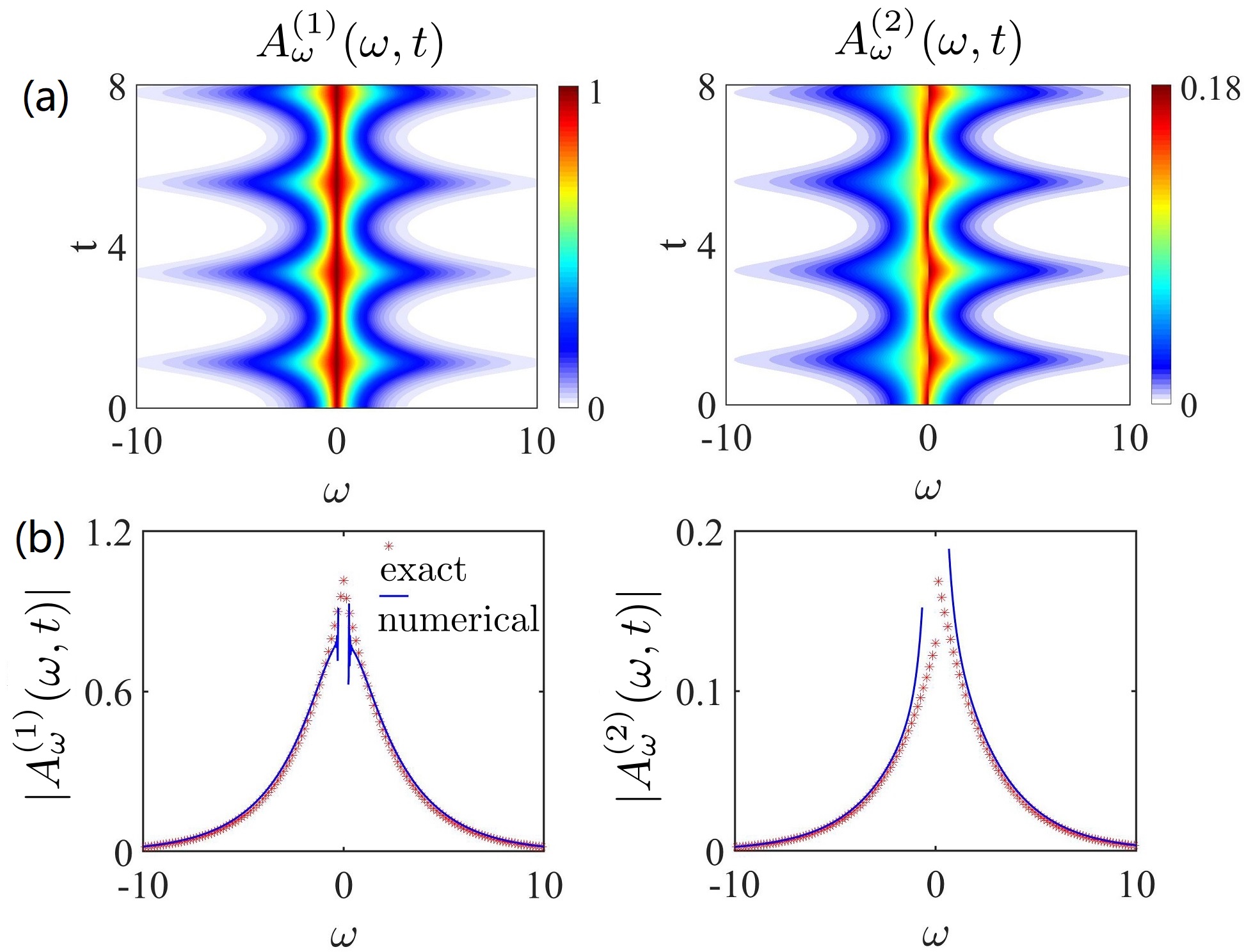}
\caption{(a) Evolution of the two spectral components $|A_\omega^{(j)}(\omega, t)|$ of vector KMS given by Eqs. (\ref{FT-KMB}) when $\omega'=\omega-\beta_j$. These spectra correspond to the KMS solution shown in Fig. \ref{f1} (c)].  (b) Comparison of numerical (solid curves) and exact (dots) data at the point of the widest spectra ($t\approx1$).}
\label{ft}
\end{figure}

Figure \ref{ft}(a) shows the spectral evolution of the vector bright-dark KMS given by Eqs. (\ref{FT-KMB}) with $\omega'=\omega-\beta_j$.
The spectra correspond to the amplitude profiles shown in Fig. \ref{f1}(c). In each component, the spectrum is periodic along the $t$-axis just as the KMS itself.
The spectra are widest at the points of the maximal selfcompression of the soliton.
We have compared the exact spectra with the numerical results obtained by the numerical integration for the wave fields factored by a super-Gaussian function. As shown in Fig. \ref{ft} (b), the spectral profiles are very close to each other.

\section{eigenvalue analysis, KMS existence diagrams and critical relative wavenumber}\label{sec4}

\begin{figure}[htbp]
\centering
\includegraphics[width=84mm]{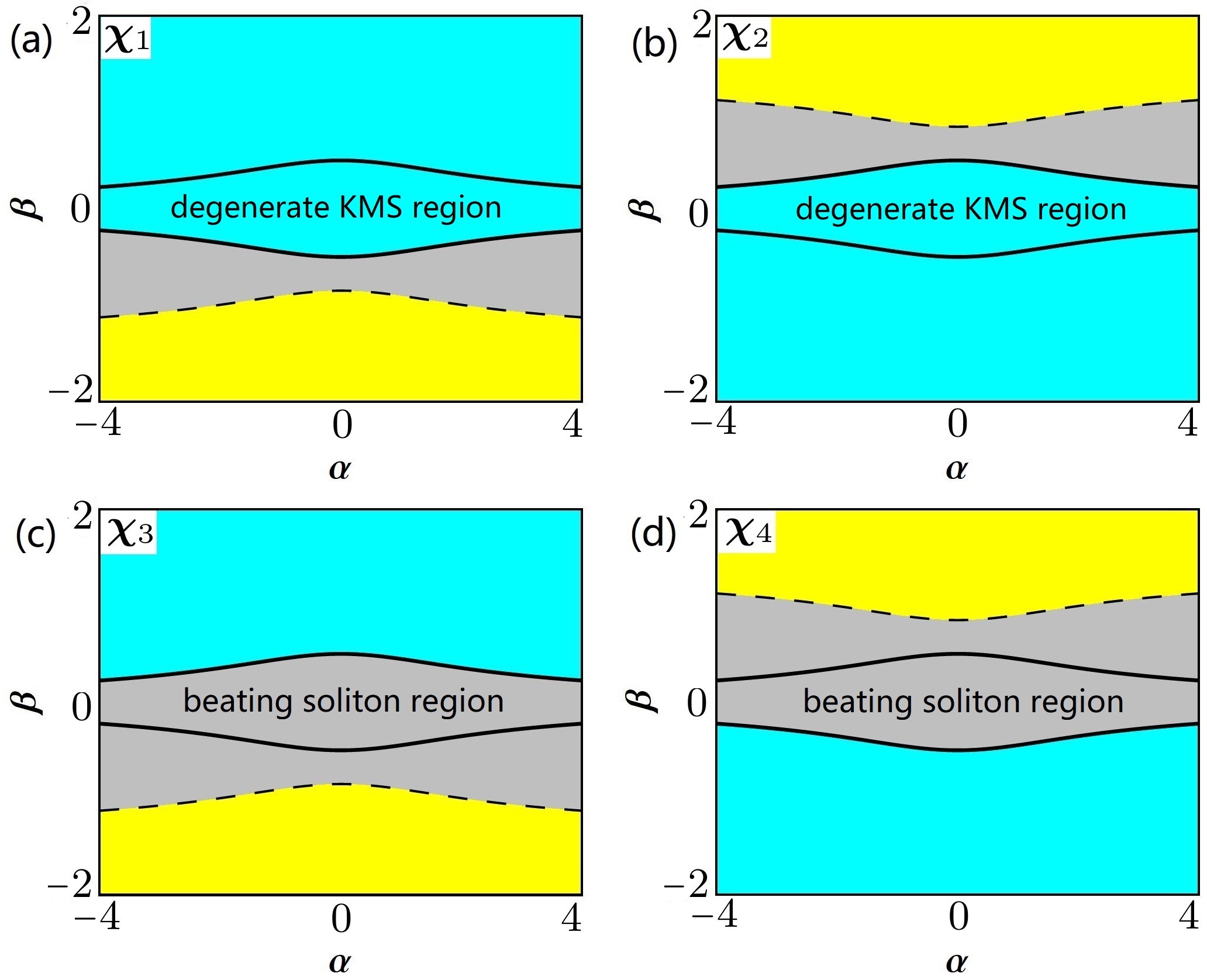}
\caption{Existence diagrams of vector KMSs with four eigenvalues (\ref{Eqchi1234}) in the ($\alpha$, $\beta$) plane. In all panels, cyan, yellow, and gray areas denote the bright, dark, and four-petal KMS, respectively. The two solid lines in each panel represent the critical condition (\ref{eqbc}). In the two upper panels (for $\bm{\chi}_1$, $\bm{\chi}_2$), KMSs are degenerate in the regions within these two lines ($\beta^2\leq\beta_c^2$). KMSs outside these areas $\beta^2>\beta_c^2$ are nondegenerate. In the two lower panels ($\bm{\chi}_3$, $\bm{\chi}_4$), beating solitons are obtained in the regions $\beta^2\leq\beta_c^2$. Outside of these areas, KMSs are the same as those in the cases $\bm{\chi}_1$, $\bm{\chi}_2$.
}\label{f-pd}
\end{figure}

All examples shown in Fig. \ref{f1} correspond to the solution (\ref{eqkmb}) with a single eigenvalue (i.e., $\bm{\chi}_1$, see below).
However, Eq. (\ref{eqchi}) admits multiple roots. The presence of several eigenvalues adds a new physics to the KMS in a Manakov system.
For simplicity, we consider only the case of equal background amplitudes $a_1=a_2=a$. Then, there are four eigenvalues:
\begin{eqnarray}\label{Eqchi1234}
\begin{split}
\bm{\chi}_{1}=-\frac{i}{2}\alpha-\sqrt{\kappa-\sqrt{\eta}},~~\bm{\chi}_{2}=-\frac{i}{2}\alpha+\sqrt{\kappa-\sqrt{\eta}},\\
\bm{\chi}_{3}=-\frac{i}{2}\alpha-\sqrt{\kappa+\sqrt{\eta}},~~\bm{\chi}_{4}=-\frac{i}{2}\alpha+\sqrt{\kappa+\sqrt{\eta}},
\end{split}
\end{eqnarray}
where
\begin{eqnarray}
\kappa&=&\beta^2-a^2-\alpha^2/4,\nonumber\\
\eta&=&a^4-4a^2\beta^2-\alpha^2\beta^2.\nonumber
\end{eqnarray}
Naturally, the solution (\ref{eqkmb}) with either of the eigenvalues (\ref{Eqchi1234}) satisfy the Manakov system (\ref{eq1}). However, not all four solutions are realistic.
Figure \ref{f-pd} shows the regions of existence of four vector KMSs with different eigenvalues on the ($\alpha$, $\beta$) plane. The pink, yellow and cyan areas on these plots correspond to dark, four-petal and `bright' KMS, respectively. These are defined by Eq. (\ref{EqH}) as described above. The dashed and solid curves separate the regions of different types of solitons. The solid curves are found analytically while the dashed curves are calculated numerically.

The analytic expressions for the solid curves can be obtained directly from Eq. (\ref{Eqchi1234}) using the condition $\eta=0$. These are given by:
\begin{eqnarray}
\beta^2=\beta_c^2=\frac{a^4}{4a^2+\alpha^2}.\label{eqbc}
\end{eqnarray}
This equation defines the \textit{critical} relative wavenumber that plays a key role in the properties of the vector KMS. It is represented by two solid lines in Figs. \ref{f-pd}. Namely, the KMSs are different in the regions $\beta^2\leq\beta_c^2$ and $\beta^2>\beta_c^2$.

\subsection{KMS in the case $\beta^2\leq\beta_c^2$}

\begin{figure}[htbp]
\centering
\includegraphics[width=84mm]{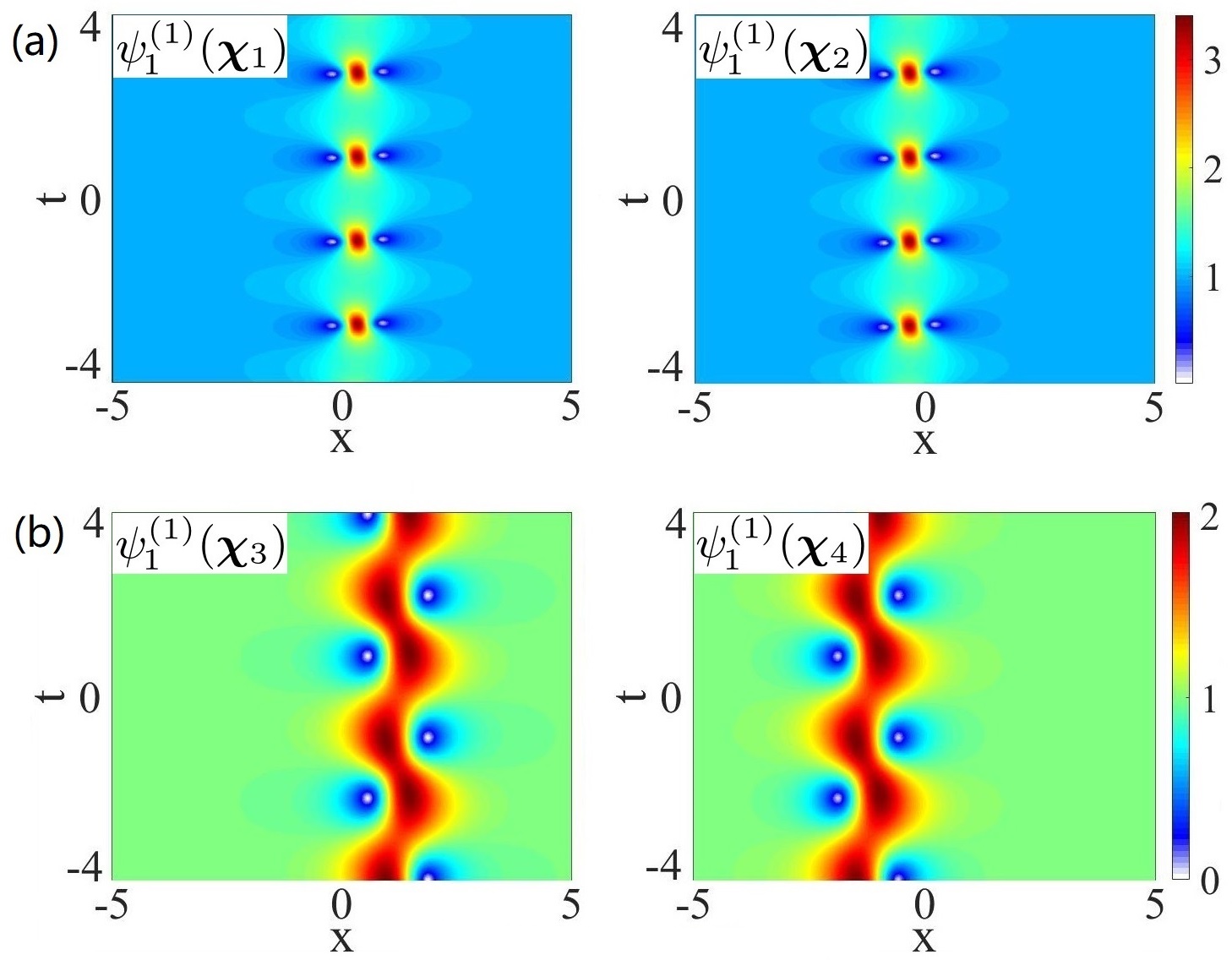}
\caption{Amplitude profiles of $|\psi_1^{(1)}|$ in the region $\beta^2\leq\beta_c^2$ with different eigenvalues (\ref{Eqchi1234}). The parameters are $a = 1$, $\alpha=2$, and $\beta=0.1$.}\label{f-chi1-4}
\end{figure}

When $\beta^2\leq\beta_c^2$, the eigenvalues (\ref{Eqchi1234}) and the corresponding Lax spectral parameters are \textit{purely imaginary}.
This means that the wave propagates along $t$ with the vanishing group velocity $V_g=0$.
Moreover, we have
\begin{eqnarray}
\bm\chi_{1i}+\bm\chi_{2i}&=&-\alpha,\\
\bm\chi_{3i}+\bm\chi_{4i}&=&-\alpha.
\end{eqnarray}
These relations satisfy the symmetry (\ref{eqsymm2}):
\begin{eqnarray}
\left|\psi^{(j)}_1[(x,t);\bm\chi_{1}]\right|&=&\left|\psi^{(j)}_1[(x',t');\bm\chi_{2}]\right|,\label{eqchi12}\\
\left|\psi^{(j)}_1[(x,t);\bm\chi_{3}]\right|&=&\left|\psi^{(j)}_1[(x',t');\bm\chi_{4}]\right|.\label{eqchi34}
\end{eqnarray}
This indicates that $\{\psi^{(j)}_1(\bm\chi_{1})$, $\psi^{(j)}_1(\bm\chi_{2})\}$ or $\{\psi^{(j)}_1(\bm\chi_{3})$, $\psi^{(j)}_1(\bm\chi_{4})\}$ have the same amplitude profiles. The only difference between them is the trivial shifts in $x$ and $t$ equal to $\Delta x$, $\Delta t$.  Figures \ref{f-chi1-4}(a) and \ref{f-chi1-4}(b) confirm this. These
figures also show that the two upper profiles $|\psi^{(j)}_1(\bm\chi_{1})|$ and $|\psi^{(j)}_1(\bm\chi_{2})|$ are the conventional bright KMS structures while the profiles $|\psi^{(j)}_1(\bm\chi_{3})|$ and $|\psi^{(j)}_1(\bm\chi_{4})|$ are the four-petal ones.

\begin{figure}[htbp]
\centering
\includegraphics[width=84mm]{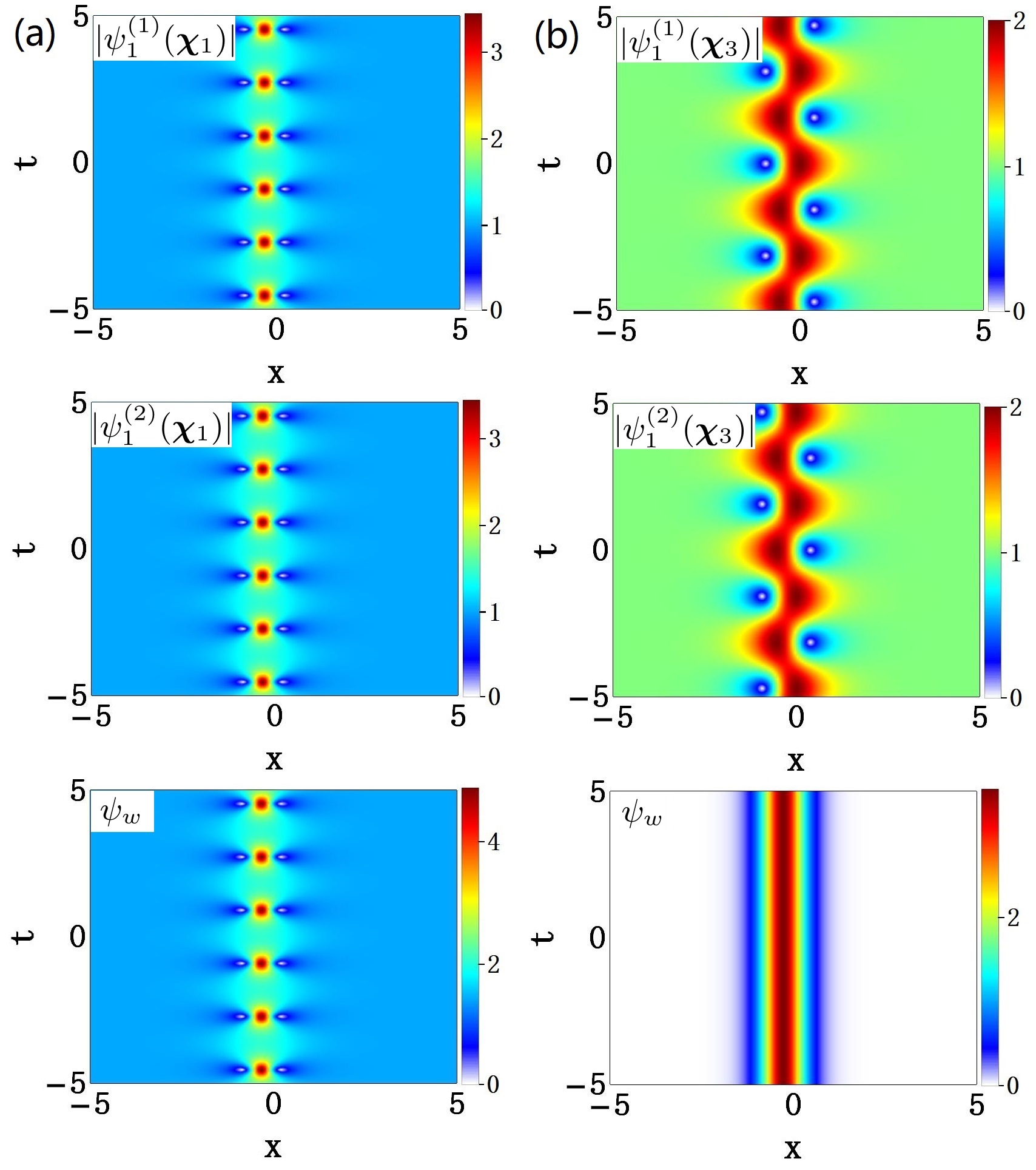}
\caption{Amplitude profiles $|\psi_1^{(j)}|$ and the total intensity distributions $\psi_w=\sqrt{|\psi_1^{(1)}|^2+|\psi_1^{(2)}|^2}$ when $\beta=0$ with the eigenvalues (a) $\bm{\chi}_1$, and (b) $\bm{\chi}_3$. Other parameters are the same as in Fig. \ref{f-chi1-4}.}\label{f-beta-0}
\end{figure}

Importantly, when $\beta^2\leq\beta_c^2$, the solutions $\psi^{(j)}_1(\bm\chi_{3})$ and $\psi^{(j)}_1(\bm\chi_{4})$ are \emph{not KMSs which are formed by the interaction between solitons and plane waves}.
In order to elucidate this point, let us consider the limit $\beta\rightarrow0$.
When the Manakov system (\ref{eq1}) decouples at $\beta\rightarrow0$, only solutions $\psi^{(j)}_1(\bm\chi_{1})$ and $\psi^{(j)}_1(\bm\chi_{2})$ become the scalar (bright) KMS [see Fig. \ref{f-chi1-4}(a)]. In the limit, $\beta=0$, we have
\begin{eqnarray}
\psi^{(1)}_1=\psi^{(2)}_1, ~~\textmd{for}~ \bm\chi_{1},~\bm\chi_{2}.\label{eqsymm1-1}
\end{eqnarray}
The relation (\ref{eqsymm1-1}) is the reduction of the vector solution to the scalar KMS.
Figure \ref{f-beta-0}(a) shows the amplitude profiles of the decoupled KMSs when the eigenvalue $\bm\chi_{1}$ is chosen.
We can see that this solution is a scalar KMS.

On the other hand, the solutions $\psi^{(j)}_1(\bm\chi_{3})$ and $\psi^{(j)}_1(\bm\chi_{4})$ have the four-petal amplitude patterns when $\beta^2\leq\beta_c^2$. Such solutions cannot be reduced to the scalar ones in the limit $\beta=0$:
\begin{eqnarray}
\psi^{(1)}_1\neq\psi^{(2)}_1, ~~\textmd{for}~ \bm\chi_{3},~\bm\chi_{4}.\label{eqsymm3-4}
\end{eqnarray}
In this limit, the solutions $\psi^{(j)}_1(\bm\chi_{3})$ and $\psi^{(j)}_1(\bm\chi_{4})$ have the form:
\begin{eqnarray}
\psi_1^{(j)}(\bm\chi_{3})&=&\psi_0^{(j)}(\psi_{DS}\mp\psi_{BS}),\\
\psi_1^{(j)}(\bm\chi_{4})&=&\psi_0^{(j)}(\tilde{\psi}_{DS}\mp\tilde{\psi}_{BS}),
\end{eqnarray}
where
\begin{eqnarray}
\tilde{\psi}_{DS}=\psi_{DS}(-x),~~\tilde{\psi}_{BS}=-\psi_{BS}(-x).
\end{eqnarray}
and
\begin{eqnarray}
\psi_{DS}=-\frac{\left(4 a^2+\alpha ^2\right) \sinh (\alpha  x)+\alpha ^2 \cosh (\alpha  x)}{\left(4 a^2+\alpha ^2\right) \cosh (\alpha  x)+\alpha ^2 \sinh (\alpha  x)},\\
\psi_{BS}=\frac{2 \left(2 a^2+\alpha ^2\right)\exp{(\frac{1}{2} i \alpha ^2 t)}}{\left(4 a^2+\alpha ^2\right) \cosh (\alpha  x)+\alpha ^2 \sinh (\alpha  x)}.
\end{eqnarray}

These solutions can be considered as \emph{linear} superpositions of the dark and bright solitons [$\psi_{DS}$, $\psi_{BS}$, $\tilde{\psi}_{DS}$, and $\tilde{\psi}_{BS}$]. These are different from the `multi-soliton complexes' which are the \emph{nonlinear} superpositions of several fundamental solitons \cite{SC1,SC2}.
Fig. \ref{f-beta-0}(b) gives an example showing  that such solution is the result of `beating effect' of vector solitons with the oscillation frequency $\alpha ^2/2$ along the $t$-axis.
The total intensity $\psi_w=\sqrt{|\psi_1^{(1)}|^2+|\psi_1^{(2)}|^2}$ shows an anti-dark soliton profile. Similar solutions can be obtained by $SU(2)$ rotations of vector dark-bright solitons  \cite{Park, EGC,LCZ}.

Thus, the solutions $\psi^{(j)}_1(\bm\chi_{3})$ and $\psi^{(j)}_1(\bm\chi_{4})$ are vector solitons in the region $\beta^2\leq\beta_c^2$
which can be interpreted as the result of linear interference between the dark and bright solitons.
Only the solutions $\psi^{(j)}_1(\bm\chi_{1})$ and $\psi^{(j)}_1(\bm\chi_{2})$ are KMSs which are formed by the interaction between solitons and plane waves when $\beta^2\leq\beta_c^2$. Moreover, for any fixed set of parameters ($a$, $\beta$, $\alpha$), the solutions $\psi^{(j)}_1(\bm\chi_{1})$ and $\psi^{(j)}_1(\bm\chi_{2})$ are degenerate KMBs.

\subsection{Non-degenerate KMS in the case $\beta^2>\beta_c^2$}

In the case $\beta^2>\beta_c^2$, all four eigenvalues are valid and satisfy the relations:
\begin{eqnarray}
&&\bm\chi_{1i}+\bm\chi_{3i}=-\alpha,~
\bm\chi_{2i}+\bm\chi_{4i}=-\alpha,\\
&&\bm\chi_{1r}=\bm\chi_{3r}=-\bm\chi_{2r}=-\bm\chi_{4r}.\label{eqchi-r}
\end{eqnarray}
The symmetry (\ref{eqsymm2}) leads to:
\begin{eqnarray}
\left|\psi^{(j)}_1[(x,t);\bm\chi_{1}]\right|&=&\left|\psi^{(j)}_1[(x',t');\bm\chi_{3}]\right|,\label{eqchi13}\\
\left|\psi^{(j)}_1[(x,t);\bm\chi_{2}]\right|&=&\left|\psi^{(j)}_1[(x',t');\bm\chi_{4}]\right|,\label{eqchi24}
\end{eqnarray}
but
\begin{eqnarray}
\left|\psi^{(j)}_1[(x,t);\bm\chi_{1}]\right|&\neq&\left|\psi^{(j)}_1[(x,t);\bm\chi_{2}]\right|.\label{eqchi12-1}
\end{eqnarray}
It follows, from Eqs. (\ref{eqchi13})-(\ref{eqchi12-1}), that for any fixed set of parameters ($a$, $\beta$, $\alpha$)  there are only two different types of vector KMSs in the region $\beta^2>\beta_c^2$. Such complexity of KMSs is absent in the scalar case.

\begin{figure}[htbp]
\centering
\includegraphics[width=84mm]{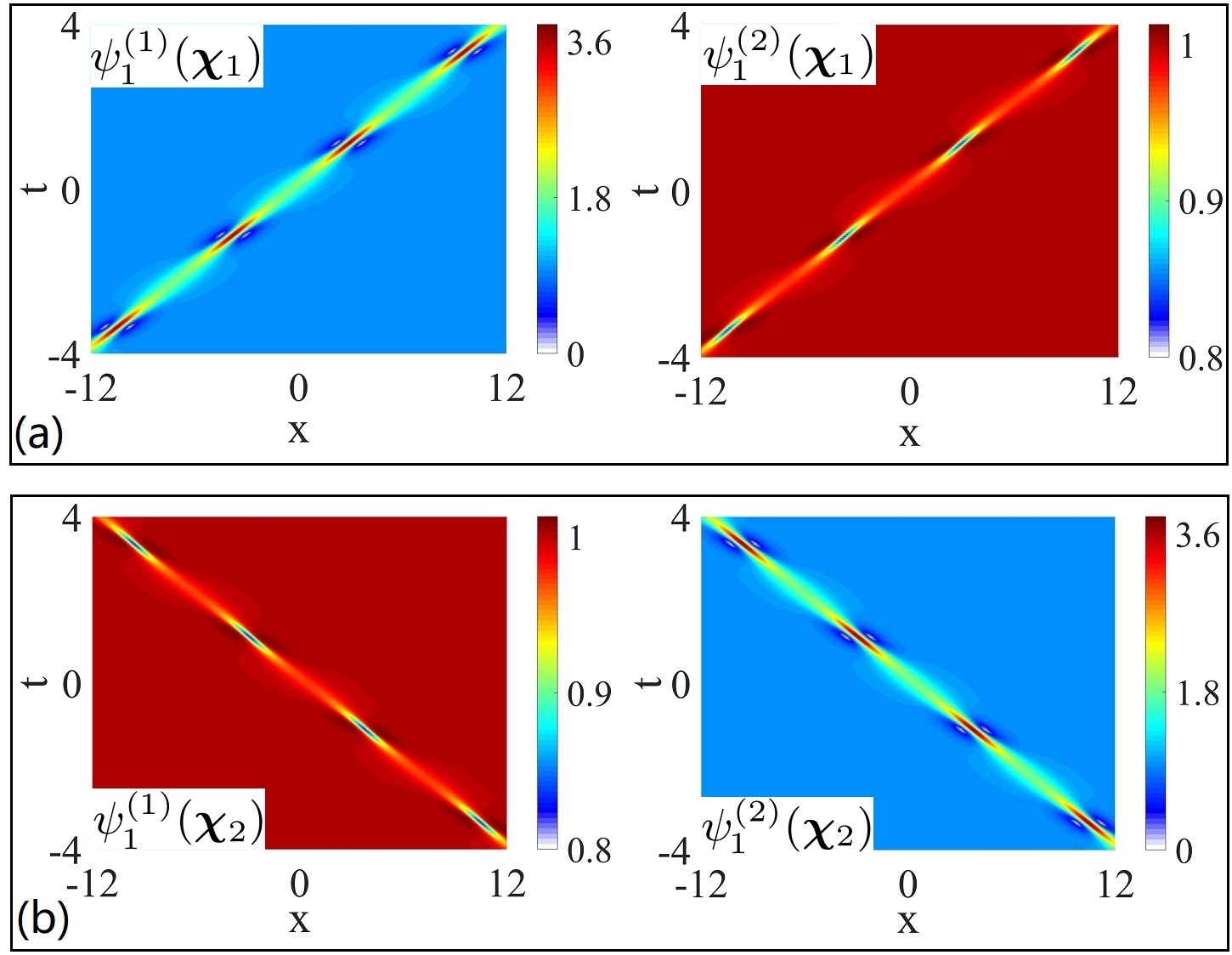}
\caption{Amplitude profiles of the vector KMSs, $|\psi^{(j)}_1|$,  with the eigenvalues (a) $\bm{\chi}_1$, and (b) $\bm{\chi}_2$, in the region $\beta^2>\beta_c^2$.
Parameters are the same as in Fig. \ref{f-chi1-4}, except for $\beta=3$.}
\label{f-nd-1}
\end{figure}

Figure \ref{f-nd-1} shows the wave profiles of KMSs $\psi^{(j)}(\bm\chi_{1})$ and $\psi^{(j)}(\bm\chi_{2})$ in the region $\beta^2>\beta_c^2$.
The first solution is a bright-dark soliton pair shown in Fig. \ref{f-nd-1}(a). It is propagating to the right with the group velocity $V_g=-\bm\chi_r$ according to Eq. (\ref{eqchi-r}). The second solution is dark-bright soliton pair shown in Fig. \ref{f-nd-1}(b). It is propagating to the left with the same group velocity $V_g=-\bm\chi_r$. They have the same period $\Omega/2\pi$ along $t$ and the same width ($\sim 1/\alpha$) in $x$.

\section{Multi-KMS in the region $\beta^2>\beta_c^2$}\label{sec5}

Each of the vector KMS can be part of the nonlinear superposition of more complex solutions.  The superposition of several KMSs can be constructed via the Darboux transformation as shown in the Appendix. In the NLSE case, such superpositions have been constructed in \cite{SKM}. Here, we concentrate on the solutions which do not have analogs in the scalar NLSE case. Namely, we consider the multi-KMSs in the region $\beta^2>\beta_c^2$ corresponding to the eigenvalues $\bm{\chi}_1$, $\bm{\chi}_2$ given by Eq. (\ref{Eqchi1234}).

\begin{figure}[htbp]
\centering
\includegraphics[width=84mm]{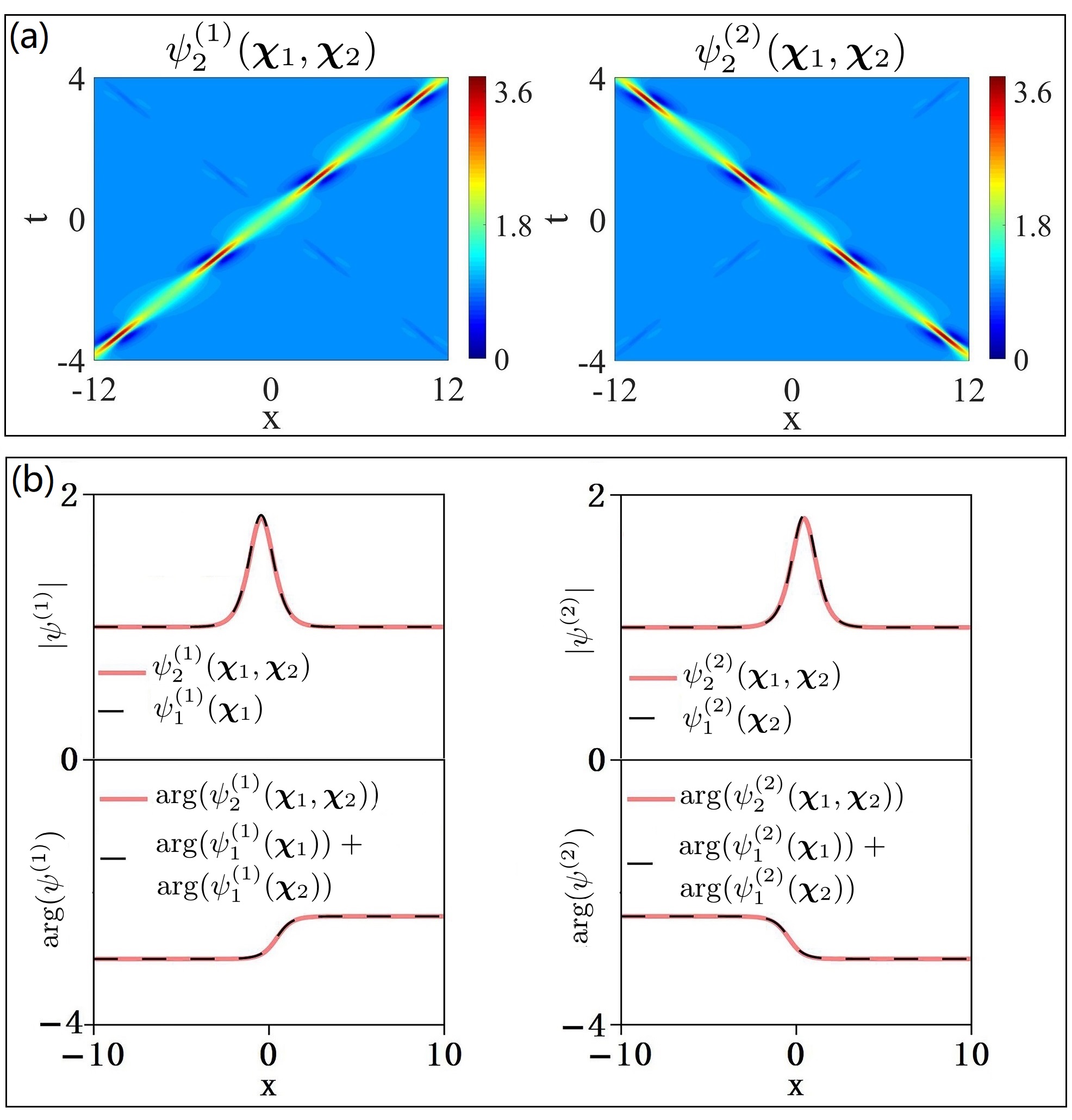}
\caption{(a) Amplitude profiles of the second-order KMSs $|\psi^{(j)}_2|$.
The plot shows the interaction of the two fundamental solitons in Fig. \ref{f-nd-1} placed on the same background.
(b) Comparison of the amplitude (upper plots) and phase (lower plots) profiles of the second order and the first order solutions  at $t=0$. Parameters are the same as in Fig. \ref{f-nd-1}.
}
\label{f-nd-2}
\end{figure}

Figure \ref{f-nd-2} shows the interaction of the two KMSs shown in Fig. \ref{f-nd-1} positioned on the same plane-wave background. The two KMSs propagate with opposite group velocities interacting at the time $t=0$. An interesting finding is that such interaction do not induce any amplitude enhancement at the point of the intersection. The two solitons pass through each other without visible mutual influence when crossing each other. This is in sharp contrast to the interaction between the scalar KMSs \cite{SKM}.

Comparison of the amplitude profiles of a single soliton and the two-soliton solution
at $t=0$ in Fig. \ref{f-nd-2} (b) shows their complete overlapping.
On the other hand, the phase jump of the two soliton interaction at $t=0$ shown in the lower part of Fig. \ref{f-nd-2} (b), is a simple sum of the phase of each KMS shown in Fig. \ref{f-nd-1}. The amplitude and phase profiles shown in Fig. \ref{f-nd-2}(b) may serve as the initial conditions for the excitation of such solution in numerical simulations.

\begin{figure}[htbp]
\centering
\includegraphics[width=84mm]{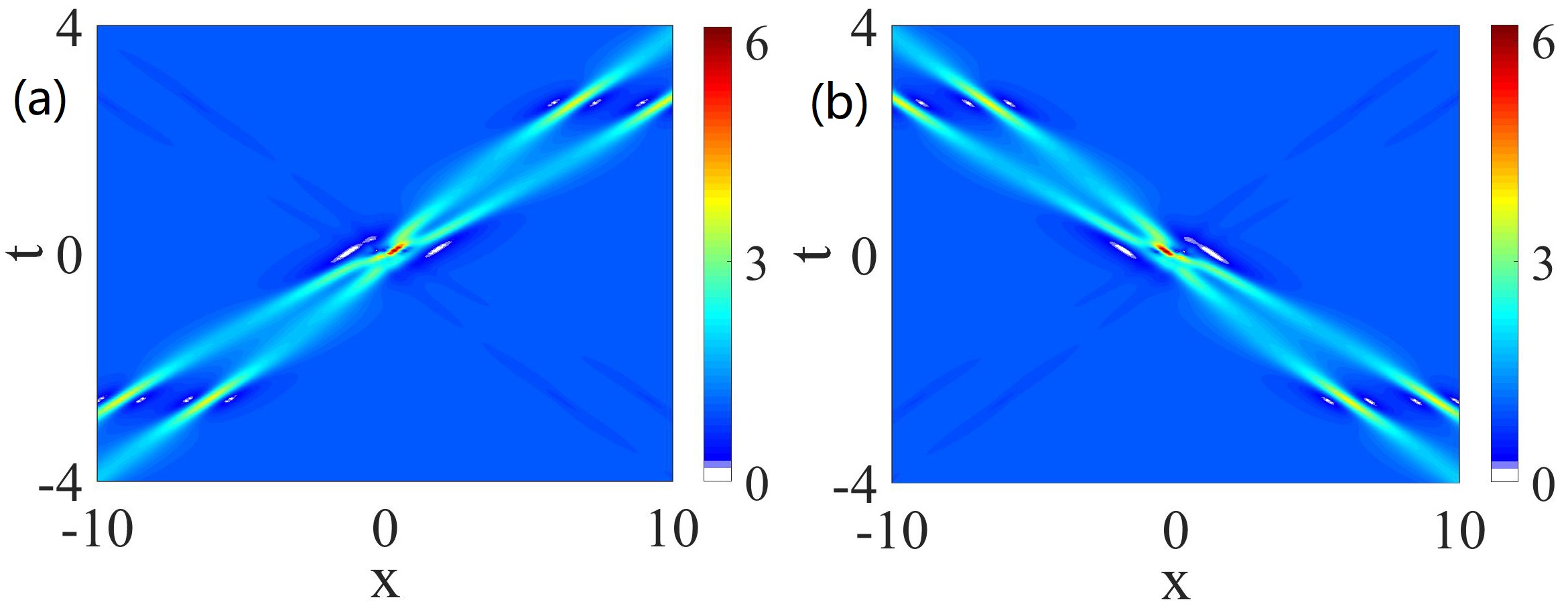}
\caption{ Amplitude profiles (a) $|\psi^{(1)}_4|$ and (b) $|\psi^{(2)}_4|$ of fourth-order solution that involves four intersecting KMSs with the eigenvalues $\bm{\chi}_1$, $\bm{\chi}_2$ given by Eq. (\ref{Eqchi1234}) and $\alpha=2$ and $\alpha=2.1$ respectively.
Other parameters are the same as in Fig. \ref{f-nd-2}.}
\label{f-nd-4}
\end{figure}

More possibilities can be realised when we consider the $2nd$-order KMS solution corresponding to the eigenvalues $\bm{\chi}_1$, $\bm{\chi}_2$ with different $\alpha$ in the region $\beta^2>\beta_c^2$.
Figure \ref{f-nd-4} shows the fourth-order KMS formed by two pairs of vector fundamental solitons corresponding to the eigenvalues $\bm{\chi}_1$, $\bm{\chi}_2$ with $\alpha=2$ and $\alpha=2.1$.
The plot shows two pairs of bright-dark KMSs in each $\psi^{(1)}$ and $\psi^{(2)}$ wave components. The group velocities of each pair of KMSs are opposite leading to the collision of the group of the KMSs at $t=0$.

\section{Numerical simulations}\label{sec6}

From an experimental point of view, an important question is what type of initial conditions can create the vector KMS that we have derived above. Clearly, our exact solution (\ref{eqkmb}) provides an ideal initial condition at any given $t$. A convenient choice is
$t=0$. Then, if we use $\psi^{(j)}(x,t=0)$ as the initial condition, both degenerate or non-degenerate KMSs can be excited.
Another possibility is to use approximations that are relatively close to the exact solution. Below, we used the following expression:
\begin{equation}\label{Eqincon}
\psi^{(j)}=\psi_{0}^{(j)}\left[1+ \mathcal{L}^{(j)}(x/w)\right],
\end{equation}
where the localised perturbation $\mathcal{L}^{(j)}(x/w)$
is either the sech- or Gaussian function with $w$ being its width. Without loss of generality, we use a Gaussian function:
\begin{equation}
\mathcal{L}^{(j)}=s^{(j)} \exp\left[-x^2/w^2\right]\exp{i\theta},
\end{equation}
where $s^{(j)}$ and $\theta$ are the amplitudes and the phase, respectively.
We chosen the width of the localised perturbation $w$ close to that of our exact solutions, $w\sim1/\alpha$.

\begin{figure}[htbp]
\centering
\includegraphics[width=84mm]{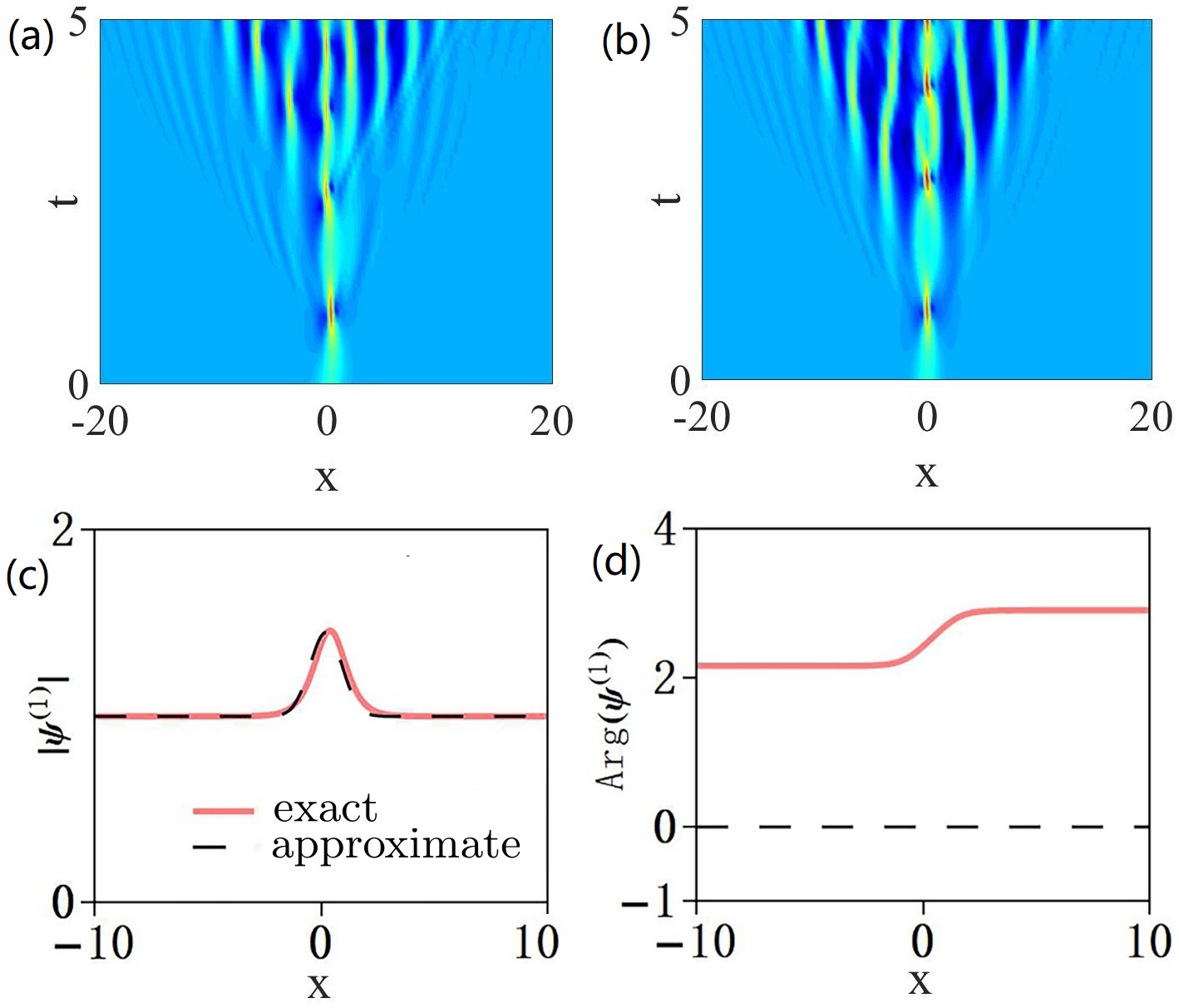}
\caption{Numerical simulations starting from (a) the exact initial condition, Eq. (\ref{eqkmb}) at $t=0$ and (b) an approximation (\ref{Eqincon}).
Parameter $\beta$ is chosen in the region $\beta^2\leq\beta_c^2$. (c) Amplitude profiles of the exact (red solid curve) and approximate (black dashed curve) initial conditions.
(d) The phase profiles of the same initial conditions. The parameters are ...}
\label{Fig-NS1}
\end{figure}

Figure \ref{Fig-NS1} depicts numerical simulations of the KMS in the first component $\psi^{(1)}$ that started with the exact initial conditions given by solution (\ref{eqkmb}) [Fig. \ref{Fig-NS1}~(a)] and the approximate initial condition (\ref{Eqincon}) [Fig. \ref{Fig-NS1}~(b)].
Simulations are done for the region $\beta^2\leq\beta_c^2$. The amplitude profiles shown in
Fig. \ref{Fig-NS1}(c) are similar in each case. However, the phase profiles shown in
Fig. \ref{Fig-NS1}(d) are different.

The fundamental KMS in each case is excited initially. However, the background is unstable and modulations around the KMS appear soon after the propagation started.
The latter is known as the auto-modulation that appears spontaneously from a localised initial modulation \cite{El,Biondini,Randoux}. Such additional modulation has been also observed in the case of the scalar NLSE \cite{MC2018}. This means that the clean observation of the KMS in experiments would be difficult. The experimental observations of the scalar KMS in an optical fibre are based on purely periodic modulation \cite{KMO}. Such technique may prevent the appearance of the the auto-modulation patterns.

\begin{figure}[htbp]
\centering
\includegraphics[width=84mm]{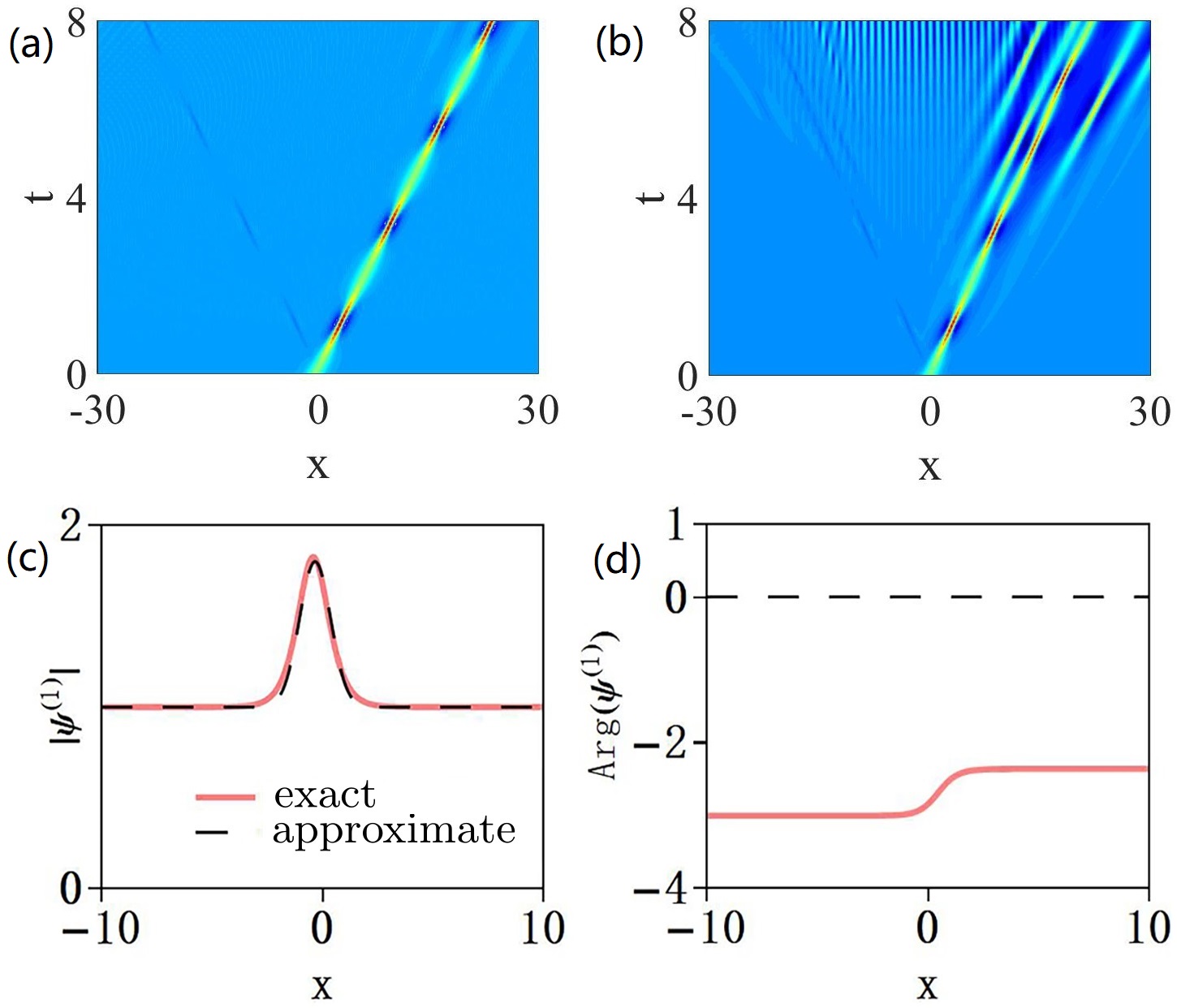}
\caption{Numerical simulations starting from (a) the exact initial condition, Eq. (\ref{eqkmb}) at $t=0$ and (b) an approximation (\ref{Eqincon}).
Parameter $\beta$ is chosen in the region $\beta^2>\beta_c^2$.
(c) Amplitude profiles of the exact (red solid curve) and approximate (black dashed curve) initial conditions. (d) The phase profiles of the same initial conditions.
 The parameters are ...}\label{Fig-NS2}
\end{figure}

Next, we consider numerical simulations of the KMS in the region $\beta^2>\beta_c^2$.
The exact and approximate initial conditions that we used here are the same as in Fig. \ref{Fig-NS1}. The results of numerical simulations of the KMS in this case are shown in Figs. \ref{Fig-NS2}(a) and \ref{Fig-NS2}(b). In each case, instead of one KMS, two KMSs propagating with opposite group velocities are excited. Numerical simulations in Fig. \ref{Fig-NS2}(a) started with the exact initial condition show a very good agreement with the exact results presented in Fig. \ref{f-nd-2}~(a). Remarkably, the auto-modulation in this case is very weak and can be seen only after four KMS periods of propagation. Approximate initial condition, on the contrary leads to quick appearance of the modulation pattern as can be seen in Fig. \ref{Fig-NS2}~(b).
 This means that accurate initial conditions provide a better way of excitation of non-degenerate KMSs in experiments.

\section{Transformation of vector KMS to an ordinary soliton}\label{sec7}

In the NLSE case, the KMS becomes a standard bright soliton at the zero amplitude of the background wave \cite{Wabnitz}. Similar transformation occurs in the case of vector KMS. This can be demonstrated directly using the exact solution (\ref{eqkmb}) by adjusting the corresponding parameters. Below, we establish the link between the vector KMS and an ordinary soliton by considering the condition of degeneracy of the eigenvalues.
Indeed, the ordinary soliton formation can be extracted from the analysis of the eigenvalues (\ref{eqchi}). Alternatively, the plain soliton solutions can be independently derived using the Darboux transformation. The details are given in the Appendix \ref{B}.

For solitons of the Manakov system (\ref{eq1}), there are two backgrounds. Therefore, the two cases can be considered separately. These are: (i) $a_1=a_2=0$ and (ii) $a_1\neq0$, $a_2=0$. We will show that in the case (i), the vector KMS is reduced to a non-degenerate bright soliton with opposite velocities of the two components. However, the case (ii) reveals a qualitatively new type of non-degenerate localised waves. Let us consider these two cases in detail.

\subsection{Non-degenerate bright solitons with $a_1=a_2=0$}\label{sec7-1}

From Eq. (\ref{Eqlambda}), we can see that the spectral parameter is
\begin{equation}\label{Eqlambda-1}
\lambda=\bm\chi.
\end{equation}
The resulting eigenvalues (\ref{Eqchi1234}) are:
\begin{eqnarray}\label{Eqchi1234-1}
\begin{split}
&\bm{\chi}_{1}=-\beta,~~\bm{\chi}_{2}=-i\alpha+\beta,\\
&\bm{\chi}_{4}=+\beta,~~\bm{\chi}_{3}=-i\alpha-\beta.
\end{split}
\end{eqnarray}
Among them, only the complex eigenvalues $\bm{\chi}_{2}$, $\bm{\chi}_{3}$ are valid.
Each of these two eigenvalues leads to the fundamental vector bright soliton.

However, a nontrivial finding is that the second-order solution with the same
eigenvalues $\bm{\chi}_{2}$ and $\bm{\chi}_{3}$ is a new family of non-degenerate bright solitons.
The derivation of these solutions is presented in Appendix \ref{B1}.
Here is the final result:
\begin{equation}\label{eqnds1}
\psi_{2}^{(j)}=\frac{\mathcal{T}\left[\left(\alpha+2i\beta\right)\cosh(\kappa_j)-\alpha\sinh(\kappa_j)\right]e^{i \phi_j}}{\mathcal{G}\cosh(2\alpha\beta t)+
\mathcal{N}\cosh(2\alpha x)-\alpha^2\sinh(2\alpha x)},
\end{equation}
where the values $\kappa_j$, $\phi_j$ are given by
\begin{equation}\label{}
\kappa_j=\alpha\left(x\pm\beta~t\right),~
\phi_j=\frac{1}{2}\left(\beta^2-\alpha^2\right)~t\pm\beta~x,\nonumber
\end{equation}
and the coefficients $\mathcal{T}$, $\mathcal{G}$ and $\mathcal{N}$ are:
\begin{equation}\label{}
\mathcal{T}=2\alpha \left(i\alpha+\beta\right),~~~
\mathcal{G}=2\left(\alpha^2+\beta^2\right),~~~
\mathcal{N}=\alpha^2+2\beta^2.\nonumber
\end{equation}

\begin{figure}[htbp]
\centering
\includegraphics[width=84mm]{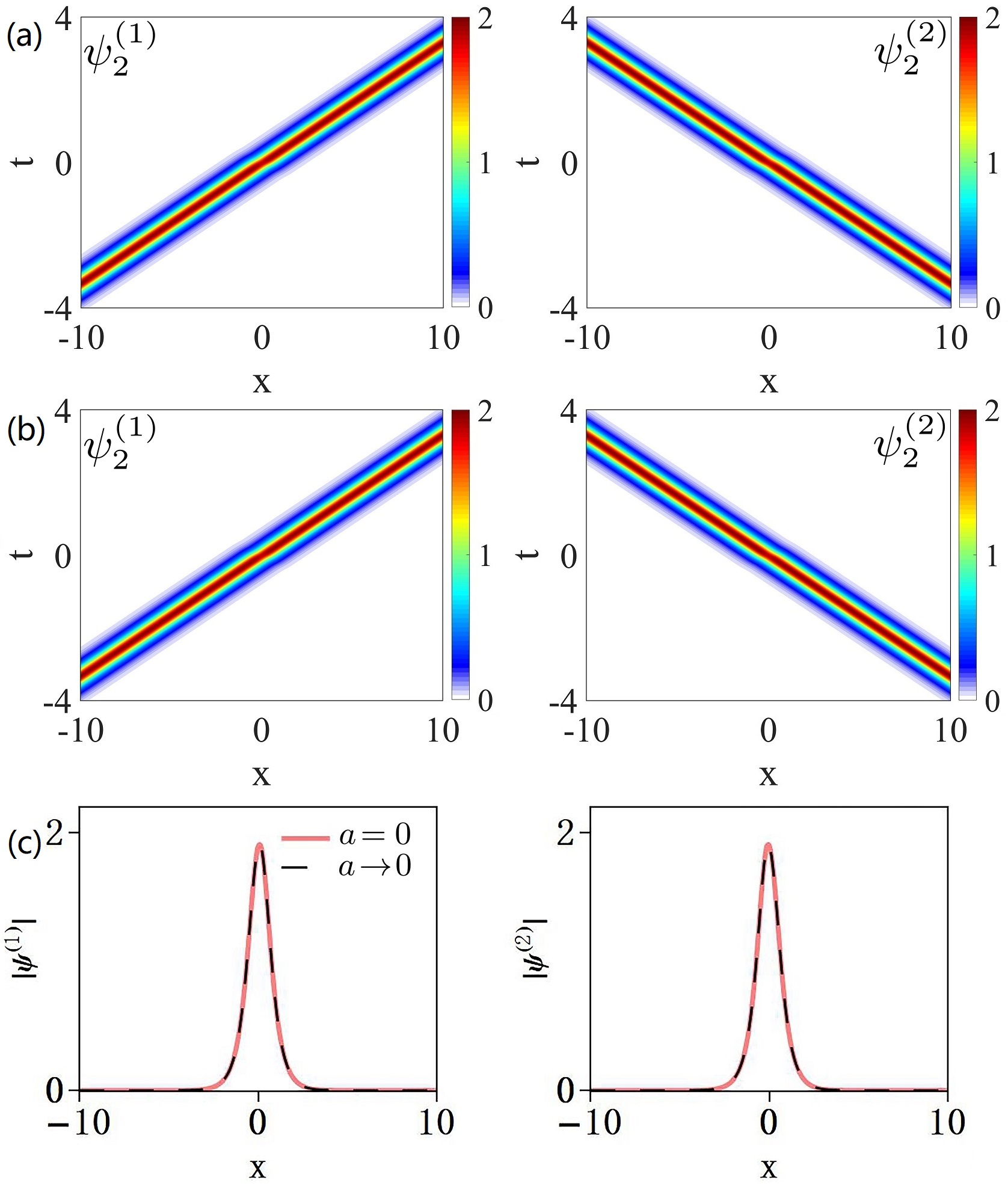}
\caption{(a) Amplitude profiles of non-degenerate soltions given by Eq. (\ref{eqnds1}), with $\beta=3$, $\alpha=2$. (b) Amplitude profiles of non-degenerate KMS in the limiting case $a_1=a_2=a\rightarrow0$. We used $a=10^{-4}$ in the solutions shown in Fig. \ref{f-nd-2}. (c)
Comparison of the soliton profiles shown in (a) and (b) at $t=0$.}\label{NDS1}
\end{figure}

Figure \ref{NDS1} (a) shows the amplitude profiles of the non-degenerate vector solitons given by Eq. (\ref{eqnds1}).
The distinctive feature of this solution is that there is only \textit{one soliton} in each wave component.
However, solitons in different wave components have opposite group velocities.
This can also be seen from the expression for $\kappa_j$ in (\ref{eqnds1}).

More detailed comparison of the second order soliton solution with a limiting case of the second order KMS is provided in Fig. \ref{NDS1}. Figure \ref{NDS1}(a) shows the non-degenerate second-order soliton solution while Fig. \ref{NDS1} (b) shows the second-order KMS solution in the limiting case of $a_1\rightarrow0$, $a_2\rightarrow0$.
This is the same solution as in Fig. \ref{f-nd-2} but in the limit of zero background.
As expected, the plots in Figs. \ref{NDS1}(a) and \ref{NDS1}(b) are identical. Comparison of the soliton profiles at the point $t=0$ confirms additionally that the two second-order solutions have the same profiles. Interestingly, there is no visible interaction between the two solitons.

More complex patterns can be revealed from the fourth-order solutions derived in Appendix \ref{B1}. Figure \ref{NDS2} (a) displays the two wave components of the fourth-order soliton solution with the eigenvalues ($\bm{\chi}_{2}$, $\bm{\chi}_{3}$) where $\alpha=2$ and $\alpha=2.1$, respectively. These patterns show the interaction between the non-degenerate solitons.
However, only two solitons interact with each other in each wave component. Again,
there is no interaction between different wave components although the two pairs of solitons have opposite velocities and collide at the point $t=0$.

\begin{figure}[htbp]
\centering
\includegraphics[width=84mm]{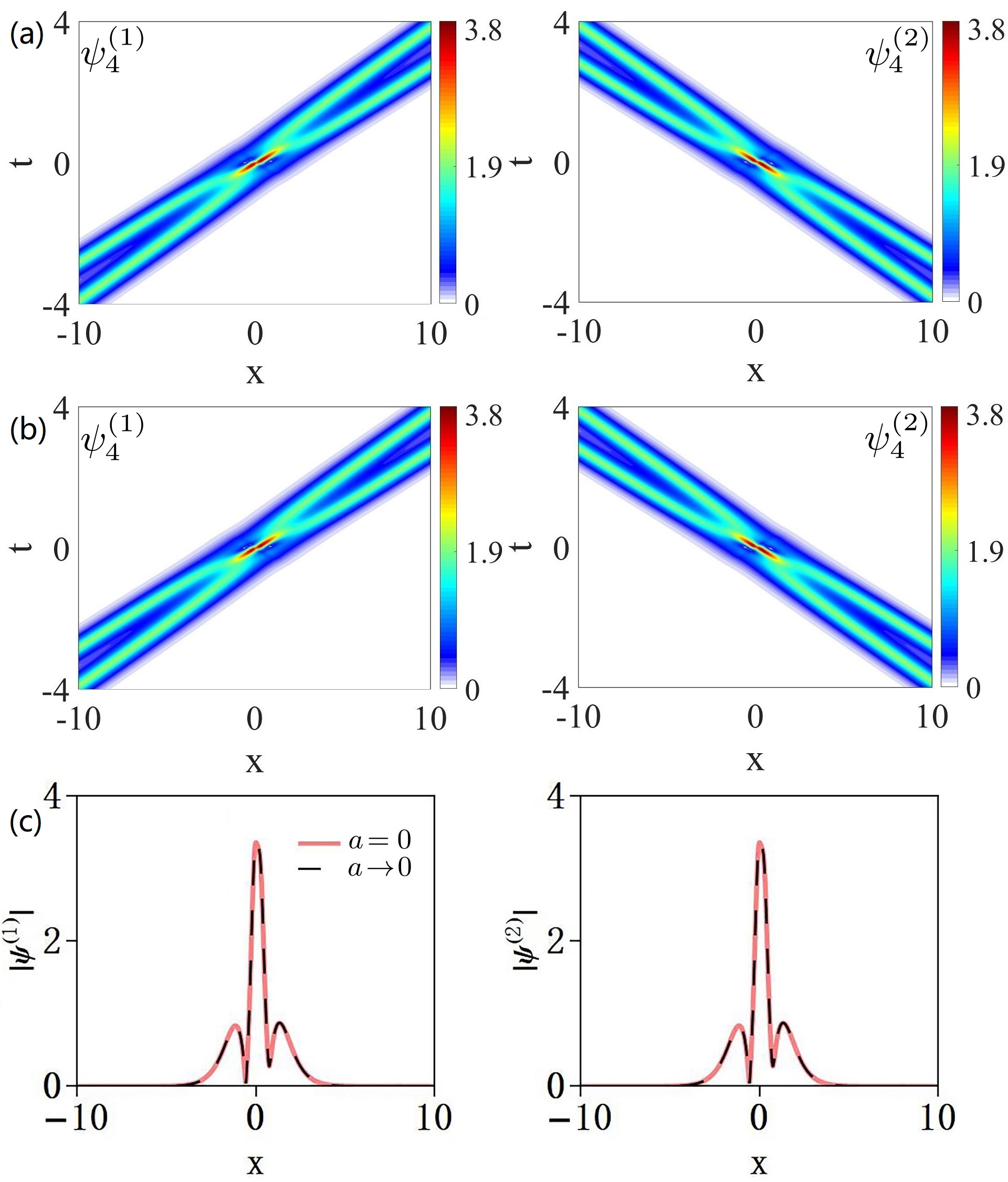}
\caption{(a) Amplitude profiles of non-degenerate fourth-order solutions given by Eqs. (\ref{Eq-NS1}) and (\ref{Eq-NS2}), with $\beta=3$, $\alpha=2$, $\alpha=2.1$. (b) Amplitude profiles of the two second-order non-degenerate KMSs. These are limiting cases of the solutions shown in Fig. \ref{f-nd-4} when $a_1=a_2=a\rightarrow0$. In order to avoid numerical artefacts, we use $a_1=a_2=10^{-4}$. (c)
Comparison of the wave profiles shown in (a) and (b) at the point $t=0$.}
\label{NDS2}
\end{figure}

The same fourth-order soliton solution can be obtained as the limiting case of the fourth-order KMS solution shown in Fig. \ref{f-nd-4} but when $a_1\rightarrow0$, $a_2\rightarrow0$. This solution is shown in Fig. \ref{NDS2} (b). The two fourth-order solutions shown in Figs. \ref{NDS2}(a) and \ref{NDS2}(b) are identical. This can also be seen from the comparison of the soliton profiles at $t=0$ shown in Fig. \ref{NDS2}(c).


\subsection{Non-degenerate localized waves with $a_1\neq0$, $a_2=0$}\label{sec7-2}
When $a_1\neq0$, $a_2=0$, the spectral parameter defined by Eq. (\ref{Eqlambda})  is:
\begin{equation}\label{Eqlambda2}
\lambda=\bm\chi-\frac{a_1^2}{\bm\chi+\beta_1}.
\end{equation}
The corresponding eigenvalues $\bm\chi$ are obtained explicitly from Eq. (\ref{eqchi}):
\begin{eqnarray}\label{Eqchi1234-2}
\begin{split}
&\bm{\chi}_{1}=\beta_1+\frac{i}{2}\alpha+i\sqrt{4a_1^2+\alpha^2},~~\bm{\chi}_{2}=-\beta_2-i\alpha,\\
&\bm{\chi}_{3}=\beta_1+\frac{i}{2}\alpha-i\sqrt{4a_1^2+\alpha^2},~~\bm{\chi}_{4}=-\beta_2.
\end{split}
\end{eqnarray}
Here, the three complex eigenvalues $\bm{\chi}_{1}$, $\bm{\chi}_{2}$, $\bm{\chi}_{3}$ are valid.

The use of only $\bm{\chi}_{1}$ or $\bm{\chi}_{3}$ as the eigenvalue in the first step of Darboux transformation leads to a bright KMS in $\psi^{(1)}$ wave component and a zero solution in $\psi^{(2)}$. This solution is given by Eq. (\ref{eqdtt}). It is the KMS solution of the nonlinear Schr\"dingier equation.
As
\begin{eqnarray}
\bm\chi_{1i}+\bm\chi_{3i}=-\alpha,~~\bm\chi_{1r}=\bm\chi_{3r}.\label{eqchi13-r}
\end{eqnarray}
the use of either of $\bm{\chi}_{1}$, $\bm{\chi}_{3}$ leads to the same result.
On the other hand, the use of $\bm{\chi}_{2}$ as the eigenvalue in the first step of Darboux transformation results in the exact solution in the form of the vector dark-bright soliton.

\begin{figure}[htbp]
\centering
\includegraphics[width=84mm]{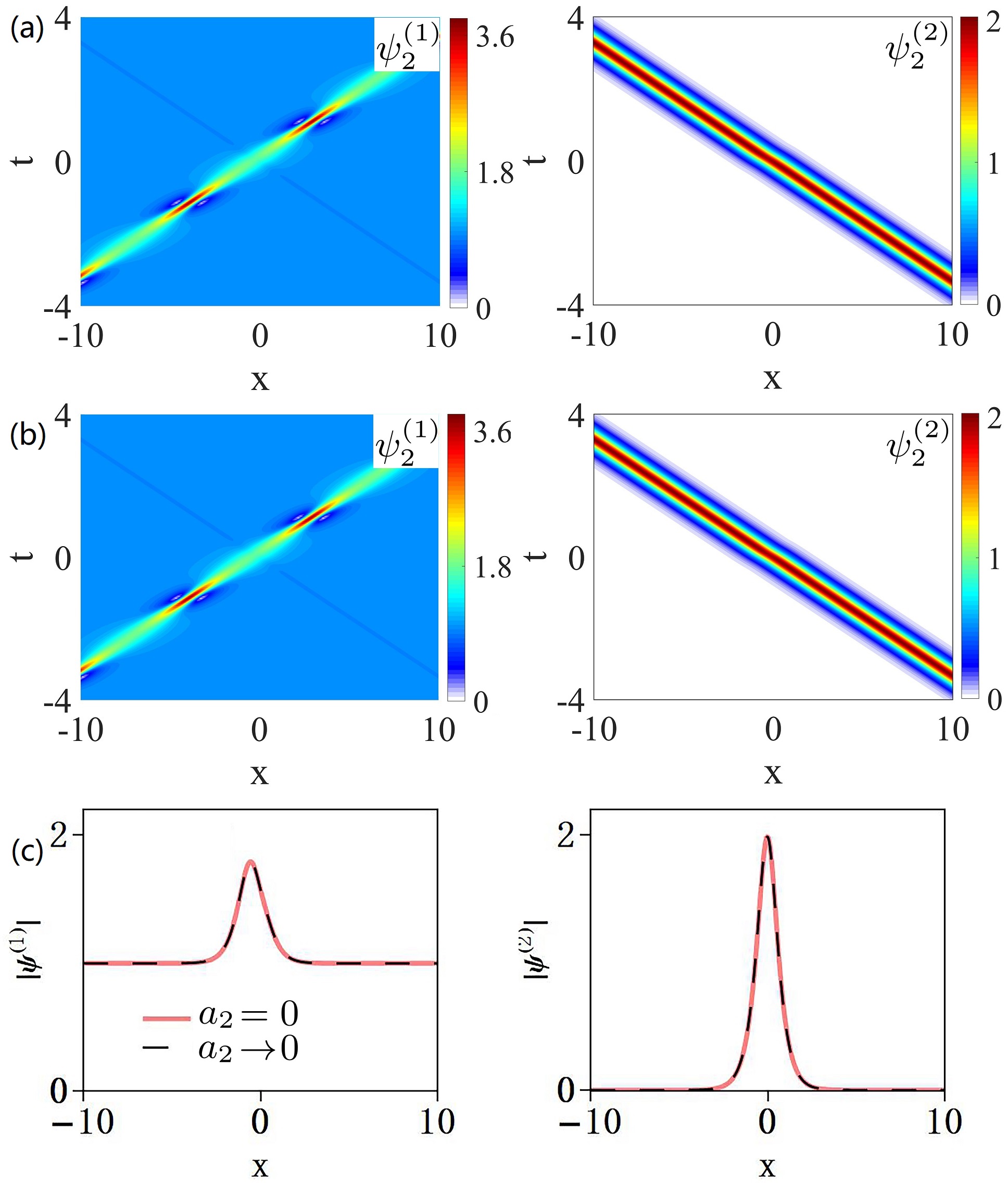}
\caption{(a) Amplitude profiles of non-degenerate soltions given by Eq. (\ref{eqnds1}), with $\beta=3$, $\alpha=2$. (b) Amplitude profiles of non-degenerate KMSs in the limiting case of $a_1=1$, $a_2\rightarrow0$ (we use here $a_2=10^{-4}$). This is a limiting case of the KMS shown in Fig. \ref{f-nd-2}. (c)
Comparison of the profiles shown in (a) and (b) at $t=0$.}\label{NDS4}
\end{figure}

The second step of Darboux transformation with the use of the eigenvalues, $\bm{\chi}_{1}$ (or $\bm{\chi}_{3}$), $\bm{\chi}_{2}$ results in more complex vector localised waves.
The corresponding exact solutions are presented in Appendix \ref{B2}.
Figure \ref{NDS4} shows examples of amplitude profiles of these solutions for particular values of $\beta$ and $\alpha$.
Figure \ref{NDS4} (a) corresponds to the KMS in the first wave component and bright soliton in the second component moving with the opposite group velocity.
The same solution can be obtained from the one shown in Fig. \ref{f-nd-2} in the limit $a_2\rightarrow0$. The corresponding amplitude profile is shown in Fig. \ref{NDS4} (b).
Naturally, the profiles shown in Figs. \ref{NDS4} (a) and Fig. \ref{NDS4} (b) are identical.
More evidence comes from the comparison of the wave profiles of the solutions shown in
(a) and (b) at $t=0$. This is shown in Fig. \ref{NDS4} (c). The two profiles completely overap.

\begin{figure}[htbp]
\centering
\includegraphics[width=84mm]{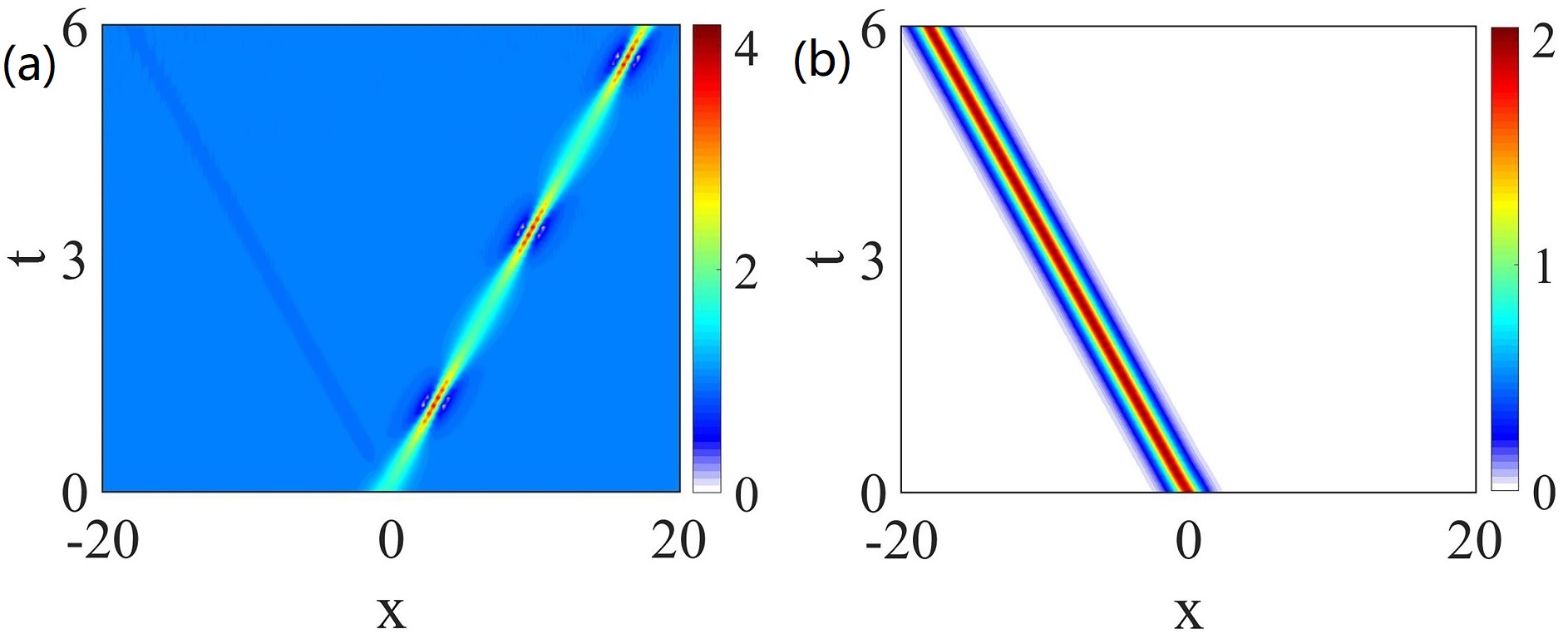}
\caption{Numerical simulations of non-degenerate waves starting with the initial condition shown in Fig. \ref{NDS4} (c) with $a_2=0$.}\label{NDS5}
\end{figure}

The wave profile shown in Fig. \ref{NDS4} (c) can be used as the initial condition for the excitation of non-degenerate waves. Such simulations will provide an independent way of proving the validity of solutions. Figure \ref{NDS5} shows the results of the simulations.
As we can see from this figure, the non-degenerate waves are well reproduced. Namely, the results are basically the same as shown in Figs. \ref{NDS4} (a) and \ref{NDS4} (b). The KMS is excited in the first component while the bright soliton with the opposite group velocity is excited in the second component.

\section{Conclusions}\label{Sec5}

In conclusion, we presented theoretical and numerical studies of vector KMSs for the Manakov equations. We derived a general family of exact vector KMS solutions of the first and higher (up to the fourth) orders that cannot be reduced to the solutions of the scalar NLSE.
Solutions that we derived can be useful for experimental works in optics, hydrodynamics and cold atom physics. One of our nontrivial findings is the prediction of a new class of non-degenerate KMSs, which has never been reported before.
They appear as higher-order solutions of the Manakov equations that form a nonlinear superposition of fundamental KMSs.
We provided the amplitude profiles for such solutions, their physical spectra, and confirm our exact solutions by numerical simulations. We also considered the limiting case of zero background when the KMS is reduced to ordinary soliton solutions. This way, we found two new families of  non-degenerate solitons.

\section*{ACKNOWLEDGEMENTS}
This work is supported by the NSFC (Grants No.
12175178, No. 12004309, No. 12022513, No. 12047502, No.
11705145, and No. 11947301),  the Major Basic Research Program of Natural Science of Shaanxi Province (No. 2017KCT-12).

\begin{appendix}
\section{Vector KMS solutions}\label{A}

We represent Eqs. (\ref{eq1}) as the condition of compatibility of two linear equations with $3\times3$ matrix operators:
\begin{eqnarray}
\bf{\Psi}_x=\textbf{U}\Psi,~~~\bf{\Psi}_t=\textbf{V}\Psi,\label{Lax-pair}
\end{eqnarray}
where $\mathbf{\Psi}= (R, S, W)^\textsf{T}$ ($\textsf{T}$ means a matrix transpose) and
\begin{eqnarray}
&&\mathbf{U}=i\left(
\begin{array}{ccc}
\lambda & \psi^{(1)*} & \psi^{(2)*}\\
\psi^{(1)}  &0 & 0\\
\psi^{(2)} & 0 &0\\
\end{array}
\right),\\
&&\mathbf{V}=i\frac{\mathbf{U}^2}{2}
\\ \nonumber
&&+i\left(
\begin{array}{ccc}
\bm{a}^2+\lambda^2& \psi^{(1)*}\lambda & \psi^{(2)*}\lambda\\
\psi^{(1)}\lambda & \bm{a}^2+|\psi^{(1)}|^2 & \psi^{(1)} \psi^{(2)*}\\
\psi^{(2)}\lambda& \psi^{(1)*} \psi^{(2)} & \bm{a}^2+|\psi^{(2)}|^2\\
\end{array}
\right),
\end{eqnarray}
where $\ast$ denotes the complex conjugate, $\lambda$ is the spectral parameter, and $\bm{a}^2=a_1^2+a_2^2$.
The Manakov system (\ref{eq1}) is equivalent to the compatibility condition
\begin{eqnarray}
\mathbf{U}_t-\mathbf{V}_x+[\mathbf{U}, \mathbf{V}]=0.
\end{eqnarray}

In order to obtain the fundamental (first-order) vector KMS solution, we start with the vector plane wave (\ref{eqpw}) $\psi_{0}^{(j)}$ as the seed solution. The corresponding spectral parameter $\lambda_{[1]}$ should satisfy the relation (\ref{Eqlambda}).
The related eigenfunctions $(R_{[1]},S_{[1]},W_{[1]})$ are given by
\begin{eqnarray}
&&R_{[1]}=\varphi_{[1]}+\tilde{\varphi}_{[1]},\\
&&S_{[1]}=\psi_{0}^{(1)}\left(\frac{\varphi_{[1]}}{\beta_1+\bm\chi_{[1]}}+\frac{\tilde{\varphi}_{[1]}}{\beta_1+\tilde{\bm\chi}_{[1]}}\right),\\
&&W_{[1]}=\psi_{0}^{(2)}\left(\frac{\varphi_{[1]}}{\beta_2+\bm\chi_{[1]}}+\frac{\tilde{\varphi}_{[1]}}{\beta_2+\tilde{\bm\chi}_{[1]}}\right),
\end{eqnarray}
where $\tilde{\bm\chi}_{[1]}=\bm\chi_{[1]}+i\alpha$, with $\bm\chi_{[1]}$ being one of the complex roots of Eq. (\ref{eqchi}).
As mentioned above, the choice of $\bm\chi_{[1]}$ is not arbitrary.
For the case $\beta^2\leq\beta_c^2$ shown in Section \ref{sec4}, we have to use $\bm\chi_{[1]}=\bm\chi_{1}$ or $\bm\chi_{[1]}=\bm\chi_{2}$.
Moreover,
\begin{eqnarray}
\varphi_{[1]}&=&\exp \bigg[ i\bm\chi_{[1]}\left(x+\frac{1}{2}\bm\chi_{[1]} t \right) \bigg],\\
\tilde{\varphi}_{[1]}&=&\exp {\bigg[ \{i\tilde{\bm\chi}_{[1]} \left(x+\frac{1}{2}\tilde{\bm\chi}_{[1]} t \right) \bigg]}.
\end{eqnarray}
The fundamental KMS solution is then obtained through the first step of the Darboux transformation:
\begin{eqnarray}\label{eqdt}
\begin{split}
&&\psi_{1}^{(1)}=\psi_{0}^{(1)}+\frac{(\lambda_{[1]}^*-\lambda_{[1]})R_{[1]}^*S_{[1]}}{|R_{[1]}|^2+|S_{[1]}|^2+|W_{[1]}|^2},\\
&&\psi_{1}^{(2)}=\psi_{0}^{(2)}+\frac{(\lambda_{[1]}^*-\lambda_{[1]})R_{[1]}^*W_{[1]}}{|R_{[1]}|^2+|S_{[1]}|^2+|W_{[1]}|^2}.
\end{split}
\end{eqnarray}
Eqs. (\ref{eqdt}) lead directly to Eqs. (\ref{eqkmb}).

Higher-order KMS solutions can be obtained via the iteration of a Darboux transformation from the fundamental KMS solution Eq. (\ref{eqdt}).
An alternative technique is based on a B\"{a}cklund transformation \cite{VB10}.
After performing the transformation, we obtain the general determinant form of the $N$th-order KMS solution:
\begin{eqnarray}
&&\psi_{N}^{(j)} =\psi_{0}^{(j)}\frac{\det(M_j)}{\det(M)},\\
&&M_j=(m^{(j)}_{[k1],[k2]})_{1\leq{k1},k2\leq{N}},\\
&&M =(m_{[k1],[k2]})_{1\leq{k1},k2\leq{N}},
\end{eqnarray}
where
\begin{eqnarray}\label{ys}
m_{[k1],[k2]}&=\frac{\varphi_{[k1]}+\varphi^*_{[k2]}}{\bm\chi^*_{[k2]}-\bm\chi_{[k1]}}+
\frac{\tilde{\varphi}_{[k1]}+\tilde{\varphi}^*_{[k2]}}{\tilde{\bm\chi}^*_{[k2]}-\tilde{\bm\chi}_{[k1]}}\nonumber\\
&+
\frac{\varphi_{[k1]}+\tilde{\varphi}^*_{[k2]}}{\tilde{\bm\chi}^*_{[k2]}-\bm\chi_{[k1]}}+
\frac{\tilde{\varphi}_{[k1]}+\varphi^*_{[k2]}}{\bm\chi^*_{[k2]}-\tilde{\bm\chi}_{[k1]}},
\end{eqnarray}
\begin{eqnarray}
m^{(j)}_{[k1],[k2]}&=\frac{\bm\chi^*_{[k2]}+\beta_j}{\bm\chi_{[k1]}+\beta_j}
\frac{\varphi_{[k1]}+\varphi^*_{[k2]}}{\bm\chi^*_{[k2]}-\bm\chi_{[k1]}}+
\frac{\tilde{\bm\chi}^*_{[k2]}+\beta_j}{\tilde{\bm\chi}_{[k1]}+\beta_j}
\frac{\tilde{\varphi}_{[k1]}+\tilde{\varphi}^*_{[k2]}}{\tilde{\bm\chi}^*_{[k2]}-\tilde{\bm\chi}_{[k1]}}\nonumber\\
&+\frac{\tilde{\bm\chi}^*_{[k2]}+\beta_j}{\bm\chi_{[k1]}+\beta_j}
\frac{\varphi_{[k1]}+\tilde{\varphi}^*_{[k2]}}{\tilde{\bm\chi}^*_{[k2]}-\bm\chi_{[k1]}}+
\frac{\bm\chi^*_{[k2]}+\beta_j}{\tilde{\bm\chi}_{[k1]}+\beta_j}
\frac{\tilde{\varphi}_{[k1]}+\varphi^*_{[k2]}}{\bm\chi^*_{[k2]}-\tilde{\bm\chi}_{[k1]}}.
\end{eqnarray}
Here, $m_{[k1],[k2]}$, and $m^{(j)}_{[k1],[k2]}$ represent the matrix elements of $M$ and $M_j$ in the $k1$-th row and $k2$-th column.
Moreover,
$\tilde{\bm\chi}_{[k1]}(\tilde{\bm\chi}_{[k2]})$=$\bm\chi_{[k1]}(\bm\chi_{[k2]})+i\alpha$ $(k1,k2=1, 2, 3, ... N)$, with $\bm\chi_{[k1]}(\bm\chi_{[k2]})$ being one of the complex roots of Eq. (\ref{eqchi}). The function $\varphi_{[k1]}(\varphi_{[k2]})$ is given by
\begin{eqnarray}
\varphi_{[k1]}=\exp\left[i\bm\chi_{[k1]}\left(x+\frac{1}{2}\bm\chi_{[k1]}t\right)\right],\\
\tilde{\varphi}_{[k1]}=\exp\left[i\tilde{\bm\chi}_{[k1]}\left(x+\frac{1}{2}\tilde{\bm\chi}_{[k1]}t\right)\right].\\
\varphi_{[k2]}=\exp\left[i\bm\chi_{[k2]}\left(x+\frac{1}{2}\bm\chi_{[k2]}t\right)\right],\\
\tilde{\varphi}_{[k2]}=\exp \left[ i\tilde{\bm\chi}_{[k2]} \left(x+\frac{1}{2}\tilde{\bm\chi}_{[k2]}t \right) \right].
\end{eqnarray}
Figures \ref{f-nd-2} and \ref{f-nd-4} show the amplitudes of the solutions in the cases $N=2$ and $N=4$ with the selected eigenvalues.

\section{Ordinary vector soliton solutions}\label{B}
Here we present the vector soliton solutions constructed by a Darboux transformation.
Two cases are considered: i) $a_1=a_2=0$; ii) $a_1\neq0$, $a_2=0$.

\subsection{Non-degenerate bright solitons with $a_1=a_2=0$}\label{B1}
When $a_1=a_2=0$, we have, from Eq. (\ref{Eqlambda}),
\begin{equation}
\lambda_{[1]}=\bm\chi_{[1]}.
\end{equation}
Eigenvalue $\bm\chi_{[1]}$ is given by Eq. (\ref{Eqchi1234-1}).
However, as mentioned above, we must have $\bm\chi_{[1]}=\bm{\chi}_{2}$ or $\bm{\chi}_{3}$.
Using such an eigenvalue (or spectral parameter) and solving the associated Lax pair with zero seed solution,
we have the eigenfunctions $\Phi_{[1]}=(R_{[1]}, S_{[1]}, W_{[1]})$
\begin{eqnarray}
&&R_{[1]}=\exp \left[ i\bm\chi_{[1]} \left(x+\frac{1}{2}\bm\chi_{[1]} t \right) \right],\nonumber\\
&&S_{[1]}=C_{s[1]},~~~
W_{[1]}=C_{w[1]}.
\end{eqnarray}
Here $C_{s[1]}$, $C_{w[1]}$ are real constants. Performing the Darboux transformation (\ref{eqdt}) with $\psi_{0}^{(j)}=0$, we obtain the fundamental vector bright soliton solution.
The higher-order iterations of the Darboux transformation lead to the non-degenerate soliton shown in Subsection \ref{sec7-1}.
The $N$th-order soliton solution can be written as:
\begin{eqnarray}
&&\psi^{(1)}_N=-(\lambda_{[N]}^*-\lambda_{[N]})\sum^{N-1}_{i=1}\mathbf{P}^{[N]}_{12},\label{Eq-NS1}\\
&&\psi^{(2)}_N=-(\lambda_{[N]}^*-\lambda_{[N]})\sum^{N-1}_{i=1}\mathbf{P}^{[N]}_{13},\label{Eq-NS2}
\end{eqnarray}
where
\begin{eqnarray}
&\mathbf{T}^{[N]}=\mathbf{I}-\frac{\lambda_{[N]}-\lambda_{[N]}^*}{\lambda-\lambda_{[N]}^*}\mathbf{P}^{[N]},\\
&\mathbf{P}^{[N]}=\frac{\Phi^{[N-1]}_{[N]}\Phi^{[N-1]\dagger}_{[N]}}{\Phi^{[N-1]\dagger}_{[N]}\Phi^{[N-1]}_{[N]}},\\
&\Phi^{[N-1]}_{[N]}=(\mathbf{T}^{[N-1]}\mathbf{T}^{[N-2]}...\mathbf{T}^{[1]}\mathbf{T}^{[0]})|_{\lambda=\lambda_{[N]}}\Phi_{[N]}.
\end{eqnarray}
Here, $\mathbf{T}^{[0]}=\mathbf{I}$ is the identity matrix. The eigenfunctions $\Phi_{[N]}=(R_{[N]}, S_{[N]}, W_{[N]})$ corresponding to $N$ different spectral parameters $\lambda_{[1]} , \lambda_{[2]} , ... , \lambda_{[N]}$ are given by:
\begin{eqnarray}
&&R_{[N]}=\exp\{i\bm\chi_{[N]}(x+\frac{1}{2}\bm\chi_{[N]} t)\},\nonumber\\
&&S_{[N]}=C_{s[N]},~~~W_{[N]}=C_{w[N]}.
\end{eqnarray}
Letting $C_{s[1]}=C_{w[2]}=1$, $C_{w[1]}=C_{s[2]}=0$, we obtained the non-degenerate soliton solution (\ref{eqnds1}) with $N=2$.
The profiles are shown in Fig. \ref{NDS1}.
Furthermore, the 4-order soliton solutions with $C_{s[3]}=C_{w[4]}=1$, $C_{w[3]}=C_{s[4]}=0$ describe the interaction between two non-degenerate solitons shown in Fig. \ref{NDS2}.

\subsection{Non-degenerate localized waves with $a_1\neq0$, $a_2=0$}\label{B2}

In this case, we have, from Eq. (\ref{Eqlambda}),
\begin{equation}
\lambda_{[1]}=\bm\chi_{[1]}-\frac{a_1^2}{\bm\chi_{[1]}+\beta_1}.
\end{equation}
Eigenvalue $\bm\chi_{[1]}$ is given by Eq. (\ref{Eqchi1234-2}).
However, as mentioned above, we must have $\bm\chi_{[1]}=\bm{\chi}_{1}(\bm{\chi}_{3})$ or $\bm{\chi}_{2}$.
Here, we first consider the eigenvalue $\bm\chi_{[1]}=\bm{\chi}_{1}$.
The corresponding eigenfunctions $(R_{[1]},S_{[1]},W_{[1]})$ are given by
\begin{eqnarray}\label{eqdtt-s}
&&R_{[1]}=\varphi_{[1]}+\tilde{\varphi}_{[1]},\\
&&S_{[1]}=\psi_{0}^{(1)}\left(\frac{\varphi_{[1]}}{\beta_1+\bm\chi_{[1]}}+\frac{\tilde{\varphi}_{[1]}}{\beta_1+\tilde{\bm\chi}_{[1]}}\right),\\
&&W_{[1]}=0,
\end{eqnarray}
where $\tilde{\bm\chi}_{[1]}=\bm\chi_{[1]}+i\alpha$, and
\begin{eqnarray}
\varphi_{[1]}&=&\exp{\{i\bm\chi_{[1]}(x+\frac{1}{2}\bm\chi_{[1]} t)\}},\\
\tilde{\varphi}_{[1]}&=&\exp{\{i\tilde{\bm\chi}_{[1]} (x+\frac{1}{2}\tilde{\bm\chi}_{[1]} t)\}}.
\end{eqnarray}
The first-order solution obtained through the Darboux transformation is:
\begin{eqnarray}\label{eqdtt}
\psi_{1}^{(1)}&=&\psi_{0}^{(1)}+\frac{(\lambda_{[1]}^*-\lambda_{[1]})R_{[1]}^*S_{[1]}}{|R_{[1]}|^2+|S_{[1]}|^2+|W_{[1]}|^2},
\\
\psi_{1}^{(2)}&=&0.
\end{eqnarray}
The solution (\ref{eqdtt}) contains a KMS but only in the $\psi_{1}^{(1)}$ component.

To obtain the non-degenerate localized waves shown in Section \ref{sec7-2}, we apply the second step of the Darboux transformation.
Note that the the second spectral parameter used here is different from that in the first step.
Namely, $\lambda_{[2]}=\bm\chi_{[2]}-\frac{a_1^2}{\bm\chi_{[2]}+\beta_1}$, where $\bm\chi_{[2]}=\bm{\chi}_{2}$.
The corresponding eigenvalues are given by
\begin{eqnarray}
R_{[2]}&=&\exp{ \left[ i\bm\chi_{[2]} \left(x+\frac{1}{2}\bm\chi_{[2]} t \right) \right]},\\
S_{[2]}&=&\psi_{0}^{(1)}\left(\frac{\exp{ \left[ i\bm\chi_{[2]} \left(x+\frac{1}{2}\bm\chi_{[2]} t \right) \right] }}{\beta_1+\bm\chi_{[2]}}\right),\\
W_{[2]}&=&\exp(i\theta_2)\left(\frac{\exp{ \left[ i\bm\chi_{[2]} \left(x+\frac{1}{2}\bm\chi_{[2]} t \right) \right] }}{\beta_2+\bm\chi_{[2]}}\right),
\end{eqnarray}
Finally, the second-order solution which describes the non-degenerate localized waves shown in Fig. \ref{NDS4} can be written as
\begin{eqnarray}
\begin{split}
&\psi_{2}^{(1)}=\psi_{1}^{(1)}+\frac{2i(\lambda_{[2]}^*-\lambda_{[2]})\Phi_{1}^* \Phi_{2}}{|\Phi_{1}|^2+|\Phi_{2}|^2+|\Phi_{3]}|^2},\\
&\psi_{2}^{(2)}=\frac{2i(\lambda_{[2]}^*-\lambda_{[2]})\Phi_{1}^* \Phi_{3}}{|\Phi_{1}|^2+|\Phi_{2}|^2+|\Phi_{3}|^2},
\end{split}
\end{eqnarray}
where
\begin{eqnarray}
&\Phi_{1}=\Delta\left[ \left(\frac{1}{\Delta}+\frac{R_{[1]}R_{[1]}^*}{\phi^2} \right)R_{[2]}+\frac{R_{[1]}S_{[1]}^*}{{\phi^2}}S_{[2]}\right],  \\
&\Phi_{2}=\Delta\left[\frac{S_{[1]}R_{[1]}^*}{{\phi^2}}R_{[2]}+ \left(\frac{1}{\Delta}+\frac{S_{[1]}S_{[1]}^*}{{\phi^2}} \right)S_{[2]}\right],  \\
&\Phi_{3}=W_{[2]}.
\end{eqnarray}
Here, $\Delta$=$\frac{\lambda_{[1]}^*-\lambda_{[1]}}{\lambda_{[2]}-\lambda_{[1]}^*}$ , and $|\phi|^2$=$|R_{[1]}|^2+|S_{[1]}|^2$.

\end{appendix}

\end{document}